\documentclass[footinbib,a4paper,aps,prA,reprint,twocolumn,preprintnumbers,amsmath,amssymb,10pt, superscriptaddress]{revtex4-1}
\usepackage{verbatim}
\usepackage{graphicx}
\usepackage{color}
\usepackage{tikz}
\usepackage[colorlinks=true,citecolor=blue,linkcolor=blue,urlcolor=blue]{hyperref}
\usepackage{braket}
\usepackage[T1]{fontenc}
\usepackage[normalem]{ulem}
\usepackage{upgreek}
\usepackage[percent]{overpic}
\usepackage{graphicx,xcolor}
\usepackage{dcolumn}
\usepackage{bm}
\usepackage[shortlabels]{enumitem}
\makeatletter
\renewcommand{\fnum@figure}{FIG. \thefigure}
\makeatother

\begin{document}

\title{Emergence of non-Abelian SU(2) invariance in Abelian frustrated fermionic ladders}

\author{Bachana Beradze}
\affiliation{Andronikashvili Institute of Physics, Tamarashvili str. 6, 0177 Tbilisi, Georgia} 
\affiliation{Ilia State University, Cholokashvili Avenue 3-5, 0162 Tbilisi, Georgia }

\author{Mikheil Tsitsishvili}%
\affiliation{The Abdus Salam International Centre for Theoretical Physics (ICTP), Strada Costiera 11, 34151 Trieste,
Italy}
\affiliation{International School for Advanced Studies (SISSA), via Bonomea 265, 34136 Trieste, Italy}

\author{Emanuele Tirrito}
\affiliation{The Abdus Salam International Centre for Theoretical Physics (ICTP), Strada Costiera 11, 34151 Trieste,
Italy}
\affiliation{Pitaevskii BEC Center, CNR-INO and Dipartimento di Fisica, Università di Trento, Via Sommarive 14, Trento, I-38123, Italy}

\author{Marcello Dalmonte}
\affiliation{The Abdus Salam International Centre for Theoretical Physics (ICTP), Strada Costiera 11, 34151 Trieste,
Italy}
\affiliation{International School for Advanced Studies (SISSA), via Bonomea 265, 34136 Trieste, Italy}

\author{Titas Chanda}
\email{titas.hri@gmail.com}
\affiliation{The Abdus Salam International Centre for Theoretical Physics (ICTP), Strada Costiera 11, 34151 Trieste,
Italy}
\affiliation{Department of Physics, Indian Institute of Technology Indore, Khandwa Road, Simrol, Indore 453552, India}

\author{Alexander Nersesyan}
\email{alex.a.nersesyan@gmail.com}
\affiliation{Andronikashvili Institute of Physics, Tamarashvili str. 6, 0177 Tbilisi, Georgia} 
\affiliation{Ilia State University, Cholokashvili Avenue 3-5, 0162 Tbilisi, Georgia }
\affiliation{The Abdus Salam International Centre for Theoretical Physics (ICTP), Strada Costiera 11, 34151 Trieste,
Italy}

\date{\today}

\begin{abstract}
We consider a system of interacting spinless fermions on a two-leg triangular ladder with $\pi/2$ magnetic flux per triangular plaquette. Microscopically, the system exhibits a U(1) symmetry corresponding to the conservation of total fermionic charge, and a discrete $\mathbb{Z}_2$ symmetry -- a product of parity transformation and chain permutation. Using bosonization, we show that, in the low-energy limit, the system is described by the quantum double-frequency sine-Gordon model. On the basis of this correspondence, a rich phase diagram of the system is obtained. It includes trivial and topological band insulators for weak interactions, separated by a Gaussian critical line, whereas at larger interactions a strongly correlated phase with spontaneously broken $\mathbb{Z}_2$ symmetry sets in, exhibiting a net charge imbalance and non-zero total current. At the intersection of the three phases, the system features a critical point with an emergent  SU(2) symmetry. This non-Abelian symmetry, absent in the microscopic description, is realized at low-energies as a combined effect of the magnetic flux, frustration, and many-body correlations. The criticality belongs to the SU(2)$_1$ Wess-Zumino-Novikov-Witten universality class. The critical point bifurcates into two Ising critical lines that separate the band insulators from the strong-coupling symmetry broken phase. We establish an analytical connection between the low-energy description of our model around the critical bifurcation point on one hand, and the Ashkin-Teller model and a weakly dimerized XXZ spin-1/2 chain on the other. We complement our field-theory understanding via tensor network simulations, providing compelling quantitative evidences of all bosonization predictions. Our findings are of interest to up-to-date cold atom experiments utilizing Rydberg dressing, that have already demonstrated correlated ladder dynamics.
\end{abstract}

\maketitle

\section{Introduction}

Experimental setups involving ultra-cold atoms, trapped in optical tweezer arrays and laser-coupled to highly excited Rydberg states, have demonstrated, in recent years, the remarkable potential for simulating strongly-correlated quantum phases of many-body systems under controllable experimental conditions~\cite{Schau2012, Jau2015, Faoro2016, Zeiher2016, Bernien2017, Barredo2018, Keesling2019}. 
Rydberg atom platforms, where large and long-lived van der Waals type interactions between Rydberg states can extend over relatively long distances (tunable even upto a few microns), provide unique opportunities to probe the many-body system at single-site levels with high experimental precision and control \cite{Browaeys2020} -- a feat that is
unattainable in conventional cold atoms in optical lattices governed by Hubbard-like physics (see e.g.,~\cite{Gross2017}).

In a typical experimental scenario of optical tweezer arrays, Rydberg states are populated with the dynamics between the Rydberg states being much faster compared to the respective atomic motion. 
Such a setup of Rydberg arrays is often described by interacting spin-1/2 models~\cite{Leseleuc2018, Scholl2022} paving the way to study frustrated magnetism in a controllable laboratory setting \cite{Glaetzle2015, Scholl2021, Ebadi2021, Semeghini2021}, and offers various fascinating phenomena both in one (1D) and two (2D) dimensional settings (see e.g., \cite{Schau2012, Pohl2010, Keesling2019, Lesleuc2019, Ebadi2021, Scholl2021, Verresen2021, Semeghini2021, Tarabunga2022, Jouini2023}).

In an alternative experimental scenario, the ground states of trapped ultra-cold atoms in optical lattices are weakly coupled to virtually populated Rydberg states -- the so-called `Rydberg dressing' -- resulting in generalized Hubbard-like systems with tunable long-range interactions \cite{Henkel2010,honer2010,Pupillo2010,Macri2014}.
The dynamics of such Rydberg dressed systems lies in the intermediate regime between the conventional Hubbard models describing ultra-cold atoms on optical lattices and the frozen Rydberg gases trapped in optical tweezers.
In 1D with a single bosonic field, Rydberg dressed systems show exotic critical behavior like cluster Luttinger liquids~\cite{Mattioli2013} and emergent supersymmetric critical transition~\cite{Dalmonte2015}. In 2D settings, these systems are associated with anomalous dynamics and glassy behavior~\cite{Angelone2016, Angelone2020}. Along with  these theoretical endeavors, the many-body dynamics of Hubbard models with long-range Rydberg-dressed interactions in 2D has been realized in a recent experiment~\cite{Guardado-Sanchez2021}.

While the aforementioned cases are mostly focused on either 1D or 2D geometries, in this work, we consider the intermediate regime between these two -- a ladder geometry that can accommodate interactions and magnetic terms possible in 2D geometries, while simultaneously being tractable by analytical and numerical methods that are suitable for 1D systems. In recent years, ladder systems with Rydberg dynamics have been subjected to various theoretical works, that includes coupled cluster Luttinger liquids and coupled supersymmetric critical transitions \cite{Tsitsishvili2022, Fromholz2022, Botzung2019}, Ising criticality by order-by-disorder mechanism~\cite{Sarkar2023}, chiral 3-state Potts criticality~\cite{Eck2023} -- among many others. 
Another interesting scenario that is being explored over the years in the ladder systems involves the effect of external (synthetic) magnetic flux~\cite{Orignac2001, Narozhny2005, Carr2006,  Miyake2013, Atala2014, Livi2016, Barbarino2016, Budich2017, Strinati2017, Barbarino2018, PhysRevX.7.031057,BERMUDEZ2018149,PhysRevB.99.125106,PhysRevB.106.045147, Huang2022, Bacciconi2023}, where vortex phases, topological properties, chiral boundary currents, topological Lifshitz transitions, etc. have been investigated. 

In the present work, we consider the scenario of an optical lattice system on a  two-leg triangular ladder geometry where ultra-cold spinless fermions
are trapped and subjected to a synthetic magnetic flux by means of Raman-assisted tunneling~\cite{jaksch2003creation, Dalibard2011, Celi2014, Galitski2019}. Moreover, the fermions can be coupled to Rydberg states (i.e., the Rydberg dressing) by off-resonant laser driving that can trigger controllable interactions between the fermions.
The system exhibits a U(1) symmetry corresponding to the conservation of total fermion number, and a discrete $\mathbb{Z}_2$ symmetry coming from the joint operation of parity transformation and chain inversion.
This minimal setting beyond 1D allows for the exploration of the interplay between the magnetic flux, interactions, and the geometrical frustrations, and the associated emerging phenomena, that can be investigated using well-established analytical and numerical techniques.  

In a recent paper~\cite{Beradze2023}, by considering the triangular ladder system with an asymmetric single-particle hopping across zigzag-like interchain links in the non-interacting limit,
the effect of the interplay between  geometric frustration and magnetic flux has been thoroughly investigated focusing on the single-particle band structures. 
Due to the breakdown of $k \to \pi - k$ particle-hole symmetry, two isolated low-energy Dirac-like excitations with different masses emerge. This leads to a sequence of Lifshitz transitions upon the variation of magnetic flux in the vicinity of the critical flux value of $\pi/2$ per triangular plaquette. The Lifshitz points for fixed chemical potential and fixed particle density belong to different universality classes. In the maximally frustrated case (i.e., the scenario of symmetric hopping across zigzag interchain links), the Dirac-like excitations at the boundary of the Brillouin zone become gapless, rendering the system  very susceptible to possible many-body interactions.

In this work, by utilizing a combination of field theoretical approaches based on bosonization~\cite{Nersesyan2004, Giamarchi2003} and numerical simulations based on tensor networks (TN)~\cite{schollwock_aop_2011, Orus_aop_2014, Silvi2019}, we extend this exploration by introducing many-body correlations among the fermions in the form of interchain nearest-neighbor interactions. Our particular interest is in the understanding of the interplay between geometrical frustration,  an external synthetic flux of $\pi/2$ per triangular plaquette, and many-body correlations.

We find that the system is described by the double-frequency sine-Gordon (DSG) model at low-energies. This enables us to predict a rich phase diagram that consists of trivial and topological band insulators at weak interactions and an insulating phase with spontaneously broken $\mathbb{Z}_2$ symmetry at strong couplings.
The analytical predictions coming from the phenomenological analysis of the DSG model are validated by large-scale TN simulations: crucially, we combine both matrix-product-state and tree tensor network simulations~\cite{schollwock_aop_2011, Orus_aop_2014, Silvi2019}, that, as we detail below, demonstrate complementary capabilities in probing different parts of the phase diagram.
We show that the trivial and topological band insulators are separated by a Gaussian critical line, that terminates with a Berezinskii-Kosterlitz-Thouless (BKT) transition to the strong-coupling symmetry-broken phase. 
This critical endpoint bifurcates into two Ising critical lines that separate the band insulator phases from the strong-coupling $\mathbb{Z}_2$ symmetry-broken insulator phase.

The most exciting physics lies in the low-energy description of the critical endpoint of the Gaussian critical line that bifurcates into two Ising critical lines. 
At this critical endpoint, the system exhibits enlarged non-Abelian SU(2) symmetry that is entirely absent in the microscopic description of the system, which only respects Abelian $\mathbb{Z}_2 \times$U(1) symmetry. 
Indeed, starting with our model of interacting spinless
fermions on a flux-ladder and applying the usual Jordan-Wigner (JW) transformation along the zigzag path (see
Fig.~\ref{fig:zigzag}) one would arrive at a lattice spin model which is not SU(2) invariant.
The emergence of non-Abelian SU(2) symmetry arises due to the combined effects of the synthetic magnetic flux, interactions, and the geometrical frustrations, taking place only in the low-energy limit.
At this limit, operators that explicitly violate this non-Abelian symmetry become \textit{irrelevant} in the renormalization group sense as their scaling dimensions are greater than the spacetime dimension of 2.
We show that this critical endpoint belongs to the SU(2)$_1$ Wess-Zumino-Novikov-Witten (WZNW) universality class~\cite{DiFrancesco1997, Nersesyan2004}. Furthermore, we draw an analytical connection of the low-energy description around this critical  bifurcation point to the Ashkin-Teller model~\cite{Delfino1998, Fabrizio2000, Kadanoff1981, Delfino2004} and a weakly dimerized XXZ spin-$1/2$ chain.

The paper is organized as follows. In Sec.~\ref{sec:system} we introduce the fermionic system on two-leg triangular ladder geometry and state its symmetry properties, and give a brief overview of the results of this work. In Sec.~\ref{sec:bosonization}, we provide the analytical low-energy description of the system using bosonization in terms of the DSG model, and predict the phase diagram, including SU(2)$_1$ WZNW bifurcation criticality, by phenomenological analysis of the DSG model. Section~\ref{sec:Ashkin-Teller} discusses the connection between the low-energy description of the SU(2)$_1$ WZNW criticality in this frustrated fermionic ladder with the Ashkin-Teller model and a weakly dimerized XXZ spin-1/2 chain. We validate the predictions of field-theoretical analytical treatments by performing numerical simulations based on tensor-network algorithms in Sec.~\ref{sec:numerics}. Finally, we conclude with Sec.~\ref{sec:conclu}.

\section{The System}
\label{sec:system}

\begin{figure}[tb]
    \includegraphics[width=\linewidth]{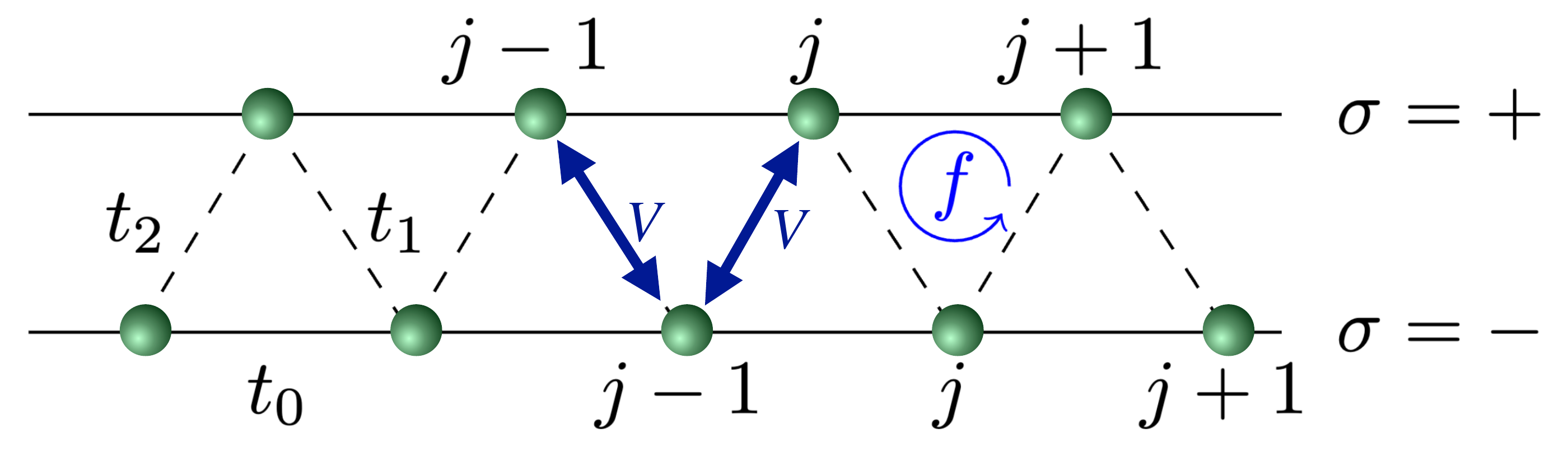}
    \caption{(Color online.) Spinless fermions on a two-leg triangular ladder. The integer $j$ labels the diatomic unit cells. $t_{0}$, $t_{1}$ and $t_{2}$ stand for the amplitudes of single-particle nearest-neighbor hopping along the chains and between them. We consider $f=1/2$ flux per triangular plaquettes (in units of $\pi$) and interchain nearest-neighbor interaction of strength $V$.
    }
    \label{fig:zigzag}
\end{figure}

We consider a paradigmatic spinless fermionic system on a two-leg triangular ladder in the presence of external magnetic flux as shown in Fig.~\ref{fig:zigzag}. Its Hamiltonian is given by
\begin{equation} \label{eq:FullHamiltonian}
    \begin{split}
         H & =H_{0} + H_{\text{int}},
        \\
         H_{0} & = -t_{0} \sum_{j,\sigma = \pm } \left(e^{-i\pi \sigma f}c^{\dag}_{j,\sigma}c^{\phantom{\dag}}_{j+1,\sigma} + \text{h.c.}\right)  
        \\
        & -\sum_{j}\left(t_{1}c^{\dag}_{j,+}c^{\phantom{\dag}}_{j,-} + t_{2}c^{\dag}_{j,+}c^{\phantom{\dag}}_{j-1,-} + \text{h.c.} \right),
        \\
        H_{\text{int}} & = V \sum_{j} \hat{n}_{j,+}  \left( \hat{n}_{j,-} +  \hat{n}_{j-1,-} \right).
    \end{split}
\end{equation}
Here $c^{\phantom{\dag}}_{j,\sigma}$ and $c_{j,\sigma}^{\dag}$ are annihilation and creation operators for a spinless fermion on the chain $\sigma = \pm$ with $j$ labeling diatomic unit cells, $\hat{n}_{j,\sigma}=c_{j,\sigma}^{\dag}c^{\phantom{\dag}}_{j,\sigma}$ are occupation number operators, $t_{0}$ and $t_{1,2}$ are  the intra- and interchain tunneling amplitudes, and $f = \Phi_{\Box}/\phi_{0}$ is the magnetic flux per triangular plaquette measured in units of the flux quantum $\phi_{0} = hc/e$. $H_{\text{int}}$ stands for the  nearest-neighbor density-density interaction of amplitude $V$ between fermions residing on top and bottom chains. We work at the regime of half-filling.

The dynamics we are interested in is relevant to cold atom gases in optical lattices, in the presence of off-resonant laser drive to Rydberg states (the Rydberg dressing)~\cite{Henkel2010,honer2010,Pupillo2010,Macri2014}. In particular, laser-dressing to $p$-states generated a strong anisotropic interaction: the latter can be made very strong vertically, and very weak horizontally, realizing the interaction pattern described by $H_{\text{int}}$~\cite{Glaetzle2015,Dalmonte2015}. Such couplings have been recently realized experimentally in Ref.~\cite{Guardado-Sanchez2021}, while distance selection (which could also be utilized for our case here) has also been demonstrated in Ref.~\cite{hollerith2022realizing}. The tunneling dynamics can instead be engineered utilizing laser-assisted tunneling~\cite{jaksch2003creation}. We note that the system is also relevant for experiments with trapped ions~\cite{Shapira23}.

\begin{figure}[htb]
    \includegraphics[width=0.8\linewidth]{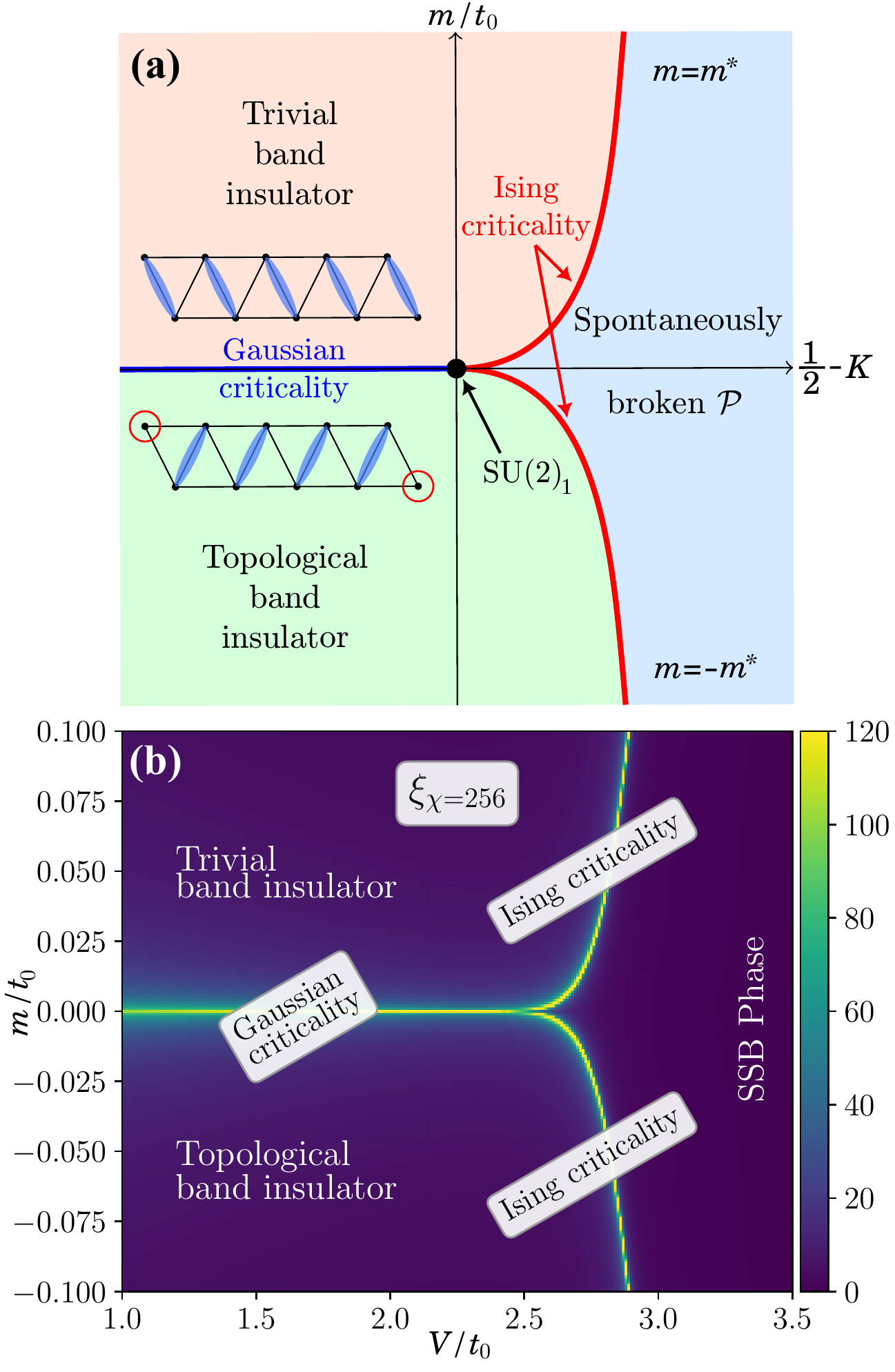}
    \caption{(Color online.) (a)  The phase diagram of the double-frequency sine-Gordon (DSG) model (Eq.~\eqref{eq:H-dsg}) that dictates that low-energy description of the lattice Hamiltonian \eqref{eq:FullHamiltonian}. The blue line corresponds to U(1) Gaussian criticality, with $c=1$. The black dot at the origin is the SU(2)$_{1}$ critical point, with $c=1$. Two red lines are the Ising criticalities, with $m= \pm m^{*}$ critical lines and $c=1/2$. In the band insulator phases, shaded in orange and green, the dimerization is non-zero along $t_{1}$ and $t_{2}$ links. The orange and green regions correspond to trivial and topological (see Appendix \ref{appendix:boundary}) band insulator phases, respectively. The region shaded in blue corresponds to the phase with spontaneously broken $\mathcal{P}$ symmetry - a combination of parity and chain interchange operations. This phase is characterized by the non-zero total current along the chains, which is proportional to the charge imbalance between the chains. 
    (b) The phase diagram of the lattice Hamiltonian \eqref{eq:FullHamiltonian} obtained by iDMRG simulations. We plot the correlation length $\xi_{\chi}$ for the iMPS bond dimension $\chi=256$ in the $(m, V/t_0)$-plane (see Sec.~\ref{sec:numerics} for details). Diverging values of the correlation length clearly indicate that the critical line at $m = 0$ bifurcates into two critical lines at around $V/t_0 \simeq 2.45$ akin to the DSG model.    
    Apart from these critical lines, all the phases are gapped, and these phases are 
    trivial and topological band insulators, and two-fold degenerate spontaneous symmetry-broken (SSB) phase.
    }
    \label{fig:Ashkin-Teller-Phase-Diagram}
\end{figure}

The properties of the non-interacting flux-ladder model $H_{0}$ have been recently studied in detail in \cite{Beradze2023}. It has been shown, that in the regime of weak interchain hopping, $0<t_{1,2} \ll t_{0}$, the effect of the geometric frustration is most pronounced in the limit $|t_{1}-t_{2}| \ll t_{1}+t_{2}$ and $f \to \frac{1}{2}$. At $f=\frac{1}{2}$ the low-energy excitations, as described by the dispersion relations
\begin{equation}
    \omega(k)_{\pm} = \pm \sqrt{4 t_0^2 \sin^2k + t_1^2 + t_2^2 + 2 t_1 t_2 \cos k} \ ,
\end{equation}
are represented by two branches of massive Dirac fermions, with momenta close to $k=0$ and $k=\pi$ in the Brillouin zone, and with masses $M=t_{1}+t_{2}$ and $m=t_{1}-t_{2}$, respectively. We will be referring to them as \textit{heavy} ($M$) and \textit{light} ($m$) fermionic sectors. The degree of frustration in the non-interacting case can be quantified by the ratio $\delta=t_{2}/t_{1}=(M-m)/(M+m)$. The model is maximally frustrated in the $m\to0$ limit. At $f=1/2$ the ground state of a half-filled ladder with $m \neq 0$ is insulating. Under the same conditions with $m = 0$, the presence of a Dirac node at $k=\pi$ renders the spectrum of the ladder gapless. Such a system is very susceptible to correlations between the particles. Consequently, we will be considering $f=\frac{1}{2}$ regime only.

The Hamiltonian Eq.~\eqref{eq:FullHamiltonian} possesses a $\mathbb{Z}_{2}$-symmetry which we label by $\mathcal{P}$: it is a product of parity transformation ($j\to-j$) and permutation of the chains ($\sigma\to-\sigma$). Obviously, $H$  has a global U(1) symmetry related to the conservation of the total particle number. However, except for the limit of two decoupled chains ($t_{1}=t_{2}=0,~f=0$), the total fermionic Hamiltonian $H$ does not display an apparent SU(2) symmetry  for any values of the parameters  of the model. One of the main results of this paper is the demonstration that, in fact, in the massless case ($m=t_{1}-t_{2}=0$) at a certain value of the coupling constant $V$ the system occurs in a critical state with central charge $c=1$, where it is characterized by the non-Abelian SU(2) symmetry.

In a symmetric flux ladder ($m=0$) at $f=1/2$, at some critical value $V_{c}$ of the interaction constant, the system undergoes a transition from a Tomonaga-Luttinger (TL) liquid phase ($V<V_{c}$) to a long-range ordered phase ($V>V_{c}$) with a spontaneously broken discrete $\mathcal{P}$-symmetry. The ordered phase is characterized by a finite interchain charge transfer and a nonzero spontaneous current along the ladder. Starting from the broken-symmetry phase and increasing the zigzag asymmetry $m$ or decreasing $V$, one observes two Ising critical lines (with central charge $c=1/2$), signifying transitions to two band insulator phases, one of them being topological ($m<0$) while the other non-topological ($m>0$). We show, both analytically and numerically, that the two Ising critical lines merge at an Ashkin-Teller (AT) bifurcation point  $V=V_{c}$, $m=0$, as shown in Fig.~\ref{fig:Ashkin-Teller-Phase-Diagram}. We show, both analytically and numerically, that at this point the symmetry of the underlying criticality is promoted to $\text{SU}(2)_{1}$ Wess-Zumino-Novikov-Witten (WZNW) universality class. The emergence of this SU(2) criticality is a remarkable property of the originally Abelian model of spinless  fermions on a triangular ladder, emerging  at low energies as a combined effect of frustration, flux, and correlations.

\section{Analytical approach to a weakly coupled flux ladder: Effective bosonized model}
\label{sec:bosonization}

In our analytical treatment of the interacting model, we concentrate on the limit of weak repulsive interaction, $V \ll W$, $W=2t_{0}$ being the ultraviolet energy cutoff. In the low-energy range, i.e., $E\sim|m| \ll M \ll W$, the most important states reside in the light sector. Interaction $H_{\text{int}}$ induces scattering processes within and between the light and heavy sectors. In the low-energy range under consideration, the interaction in the heavy sector is of minor importance, because the finite mass $M$ cuts off infrared divergences of the scattering amplitudes. Integrating out the heavy modes reduces to renormalization of the parameters of the effective Hamiltonian of the light sector. Assuming that all these renormalizations are taken into account, in what follows we will concentrate on the fermionic modes with momenta $k \sim \pi$ and small mass gap $m$.

We define the continuum limit for the lattice fermionic operators by using the correspondence
\begin{equation}
    \label{eq:c-psi}
        c_{j\sigma}\to\sqrt{a_{0}}(-1)^{j}\Psi_{\sigma}(x),
        \quad(\sigma=\pm)
    \end{equation}
where $\Psi_{\sigma}(x)$ are slowly varying fermionic fields describing single-particle excitations with momenta close to the zone boundary $k=\pi$, and $a_0$ is the lattice
constant along the chain which we set to 1. Accordingly, in the light sector, the unperturbed Hamiltonian density of the light fermionic modes takes the following continuum form
\begin{equation}
\label{eq:psi}
\begin{split}
    \mathcal{H}_{0}(x)=\Psi^{\dag}(x)
        \left[
            -\mathrm{i}v_{\text{F}}
            \left(\hat{\sigma}_{3}+\tau\hat{\sigma}_{2}\right)+
            m\hat{\sigma}_{1}
        \right]\Psi(x),
    \\ \text{with }
    \Psi=\begin{pmatrix}
            \Psi_+ \\
            \Psi_- 
        \end{pmatrix},
\end{split}
\end{equation}
where $\tau={t_{2}}/{W}=(M-m)/{2W}$ is proportional to the frustration parameter $\delta$. The kinetic energy in Eq.~\eqref{eq:psi} is brought to a canonical Dirac form by an SU(2) rotation of the spinor $\Psi$ around the $\hat{\sigma}_{1}$-axis:
\begin{equation}
\label{eq:u-transf}
\begin{split}
    \Psi(x)=U\chi(x), \hspace{10pt}
    \chi=\begin{pmatrix}
            R\\
            L
        \end{pmatrix}, \hspace{10pt} U=u+\mathrm{i}v\hat{\sigma}_{1}
    \\
    u^{2}-v^{2}\equiv\cos\gamma=\frac{1}{\sqrt{1+\tau^{2}}},
    \\
    2uv\equiv\sin\gamma=\frac{\tau}{\sqrt{1+\tau^{2}}},
\end{split}
\end{equation}
with $u^{2}+v^{2}=1$. As a consequence, in the rotated (band) basis the Hamiltonian $\mathcal{H}_{0}$ becomes
\begin{eqnarray}
\label{eq:1chi}
    \mathcal{H}_{0} (x)
        =\chi^{\dag}(x)
        \left(
            -\mathrm{i}\tilde{v}\hat{\sigma}_{3}\partial_{x}-
            m\hat{\sigma}_{1}
        \right)\chi(x),
\end{eqnarray}
where $\tilde{v}=v_{\text{F}}\sqrt{1+\tau^{2}}$ is the renormalized velocity. It is important to realize that the role of the frustration parameter $\tau$ is not exhausted by the above velocity renormalization of the single-particle excitations. As we show below, in the continuum limit, frustration in the $\tau^{2}V$-order generates pair-hopping scattering processes which, in the rotated basis, are responsible for the onset of a strong-coupling phase with broken $\mathcal{P}$  symmetry.
\medskip

In Appendix~\ref{appendix:ops}, we provide the expressions for the particle densities on each chain valid in the continuum limit in both chain and band representations. Using this expression and neglecting perturbative corrections to the frustration parameter $\tau$ (being of the order $g^n \tau$, $n\geq 1$), we obtain the continuum version of $H_{\rm int}$:
\begin{equation}
\label{eq:int-cont}
    H_{\text{int}}
        =\int\mathrm{d}x~
        \left\{
            \lambda\left(:J^{2}_{\text{R}}:+:J^{2}_{\text{L}}:\right)+
            2gJ_{\text{R}}J_{\text{L}}+\lambda\mathcal{O}_{\text{ph}}
        \right\}.
\end{equation}
Here 
\begin{equation}
\label{eq:pair-hop}
    \mathcal{O}_{\text{ph}}(x)=
        :\left(R^{\dag}L\right)_{x}
        \left(R^{\dag}L\right)_{x+a}:+\text{h.c.}
\end{equation}
is the interband pair-hopping operator, $J_{\text{R},\text{L}}(x)$ are the chiral (right and left) components of the particle density -- the U(1) chiral currents -- see Appendix \ref{appendix:ops},
\begin{equation}
\label{eq:couplings}
    \lambda=\frac{\tau^{2}g}{2(1+\tau^{2})},
\end{equation}
and $g=Va_{0}$ is a coupling constant ($a_{0}$ being the lattice constant along the chains). Recasting the kinetic energy of the fermions as a quadratic form of the chiral currents \cite{Nersesyan2004}, we arrive at the following effective continuum model describing interacting fermions in the rotated basis of states:
\begin{equation}
    \begin{split}
        \label{eq:H-J}
        \mathcal{H}(x)
        &=\mathcal{H}_{0}(x)+\mathcal{H}_{\text{int}}(x)
        \\
        &=\pi v^{*}
        \left[:J^{2}_{\text{R}}(x):+:J^{2}_{\text{L}}(x):\right] 
        \\
        &-
        mB(x)+
        2gJ_{\text{R}}(x)J_{\text{L}}(x)  +\lambda\mathcal{O}_{\text{ph}}(x),
    \end{split}
\end{equation}
where the Dirac-mass operator $B(x)$  is defined in Eq.~\eqref{eq:B-field-blu} and Eq.~\eqref{eq:D-rot}, and $v^{*}=\tilde{v}+\lambda/\pi$ is the Fermi velocity with an extra renormalization caused by interactions. Using the transformation properties of the fermionic fields, chiral currents, and mass bilinears under $\mathcal{P}$ (see Appendix  \ref{appendix:ops}, Eqs.~\eqref{eq:P-transf}), we find that $[\mathcal{H},\mathcal{P}]=0$.
\medskip

Bosonization of the model Eq.~\eqref{eq:H-J} is based on the well-know Fermi-Bose correspondence \cite{Nersesyan2004, Giamarchi2003}:
\begin{align}
\nonumber
    J_{\text{R},\text{L}}(x)
        =\frac{1}{\sqrt{\pi}}
        \partial_{x}\varphi_{\text{R},\text{L}} (x),&\quad
    \varphi_{\text{R},\text{L}}(x)
        =\frac{1}{2}\left[\pm\Phi(x)+\Theta(x)\right],\\
\label{eq:bos-rules}
    R^{\dag} (x) L(x) &= - \frac{\mathrm{i}}{2\pi \alpha} e^{- i \sqrt{4\pi} \Phi(x)}, 
\end{align}
where $\varphi_{\text{R},\text{L}}$ are chiral bosonic fields, $\Phi(x)$ and $\Theta(x)$ are the scalar field and its dual counterpart, and $\alpha \sim a_{0}$ is the short-distance cutoff of the bosonic theory. The fields $\Phi$ and $\Theta$ can be conveniently rescaled, $\Phi \to \sqrt{K} \Phi $, $\Theta\to\Theta/\sqrt{K},$ where
\begin{equation}
\label{eq:K}
    K=1-\frac{g}{2\pi v^{*}}+O(g^{2})
\end{equation}
is the so-called Luttinger-liquid interaction parameter. It decreases upon increasing the interchain repulsion $g$; however, its parametrization Eq.~\eqref{eq:K} is only universal at small values of $g$. 
Since the model at hand is not integrable, the exact analytical expression of $K=K(g)$ beyond the weak coupling limit is not known. We need to rely on numerical tools to get the dependence of $K$ at large values of $g$ (see Sec.~\ref{subsec:bifur}). Nevertheless, below we treat $K$ as an independent phenomenological parameter of the model. Collecting all the terms we arrive at the fully bosonized effective Hamiltonian which has the structure of the double-frequency sine-Gordon (DSG) model~\cite{Delfino1998, Fabrizio2000}:
\begin{equation}
\label{eq:H-dsg}
    \begin{split}
        \mathcal{H}(x)
         = & \frac{v_{c}}{2}
        \left[
            :\Pi^{2}(x):+:(\partial_{x}\Phi(x))^{2}: 
        \right]
        \\
        & +
        \left(\frac{m}{\pi \alpha}\right):\sin\sqrt{4\pi K}\Phi(x): 
        \\
        & -
        \frac{\lambda}{2(\pi\alpha)^{2}}:\cos\sqrt{16\pi K}\Phi(x):,
    \end{split}
\end{equation}
where $\Pi(x)=\partial_{x}\Theta(x)$ is the momentum canonically conjugate to the field $\Phi(x)$, and $v_{c}=v^{*}\left[1+O(g^{2})\right]$. The first two terms in Eq.~\eqref{eq:H-dsg} represent the Gaussian part of the Hamiltonian, while the remaining nonlinear terms contribute to the potential $\mathcal{U}[\Phi]$ whose profile is determined by the relative strength and signs of the $\lambda$ and $m$-perturbations. In a strong-coupling regime, the field $\Phi$ gets localized in one of the infinitely degenerate vacua of $\mathcal{U}[\Phi]$ thus determining the phase of the system. It is to be noted that when, in addition to $V$, the interaction also includes nearest-neighbor coupling along the chain -- $V_{0} \hat{n}_{j,\sigma} \hat{n}_{j+1,\sigma}$ -- the DSG model \eqref{eq:H-dsg} maintains its structure with a slightly modified velocity $v_{c}$ and the parameter $\lambda$
replaced by $\lambda = \tau^{2} (g - g_{0})/2 (1+\tau^{2})$, where $g_{0} = V_{0} a_{0}$.

Using Eqs.~\eqref{eq:ro+rot}-\eqref{eq:D-rot} and  the rules Eq.~\eqref{eq:bos-rules}, we derive the bosonized expressions of the local physical operators which will be used when discussing the correlation effects:
\small
\begin{align}
\label{tot-den}
    \rho_{\rm tot}(x)
        &=\sum_{\sigma}\rho_{\sigma}(x)=
        \sqrt{\frac{K}{\pi}}\partial_{x}\Phi(x),\\
\nonumber
    \rho_\text{rel}(x)
        &=\sum_{\sigma}\sigma\rho_{\sigma}(x)
        =\frac{1}{v_{\text{F}}} j_{0}(x),\\
\nonumber
        &=-\frac{1}{\sqrt{1+\tau^{2}}}
        \Big[
            \frac{1}{\sqrt{\pi K}}\partial_{x}\Theta(x)\\
\label{rel-den}
            &\hspace{25mm}
            -\left(\frac{\tau}{\pi\alpha}\right)
            :\cos\sqrt{4\pi K}\Phi(x):
        \Big],\\
\nonumber
    j_{z}(x)
        &=-\frac{v_{\text{F}}}{\sqrt{1+\tau^{2}}}
        \Big[
            \frac{\tau^{2}}{\sqrt{\pi K}}\partial_{x}\Theta(x)\\
\label{j1-j2-bos}
            &\hspace{25mm}
            +\left(\frac{\tau}{\pi\alpha}\right)
            :\cos\sqrt{4\pi K}\Phi(x):
        \Big],\\
\label{D1-2-bos}
    B(x)&=-\frac{1}{\pi \alpha}:\sin\sqrt{4\pi K}\Phi(x):.
\end{align}
\normalsize

\subsection{Correlation effects}

The relevance of the two operators entering the nonlinear potential $\mathcal{U}(\Phi)$  of the DSG model \eqref{eq:H-dsg} is determined by their scaling dimensions: $d_{m}=K$ and $d_{\lambda}=4K$. We will be mainly concerned with the case of a repulsive interchain interaction, $g>0$, $K<1$, and briefly comment on the attractive case $g<0$, $K>1$. At $K>1/2$ the $\lambda$-term in (\ref{eq:H-dsg}), which describes interband pair-hopping processes, is irrelevant, and the properties of the model are determined by the single-particle mass perturbation. It is well known \cite{Giamarchi2003} that, in one-dimensional models with short-range repulsive interactions, increasing local repulsion between the particles to push $K$ to small enough values may not be enough; longer-range interaction should be also invoked. Therefore, we will phenomenologically assume that interaction in the model is generalized in such a way that the regime with $K < 1/2$, where the $\lambda$-perturbation becomes relevant, is feasible. In this case, the DSG model (\ref{eq:H-dsg}) describes the interplay of correlations and single-particle perturbations. Below we discuss the possible realization of different ground-state phases of the system.

\subsubsection{$m = 0$ regime}
\label{sec:m0}

This is the case of a symmetric triangular flux ladder ($t_{1}=t_{2}$, i.e., $m=0$) in which bare fermions with momenta $k\sim \pi$ are massless. The effective  Hamiltonian Eq.~\eqref{eq:H-dsg} reduces to a standard  sine-Gordon (SG) model:
\begin{equation}
\label{eq:SG11}
    \begin{split}
        \mathcal{H} (x) &= \frac{v_{c}}{2}
            \left[
                \Pi^{2} (x) + 
                ( \partial_{x} \Phi (x) )^{2}
            \right]
        \\
         & - \frac{\lambda}{2(\pi\alpha)^{2}}
         :\cos \sqrt{16 \pi K} \Phi (x):.
    \end{split}
\end{equation}

When interchain repulsion $V$ is not strong enough and $K > 1/2$, the $\lambda$-perturbation is irrelevant, and in the infrared limit the  Hamiltonian Eq.~\eqref{eq:SG11} flows to a Gaussian model. The latter describes a Tomonaga-Luttinger liquid phase with a gapless spectrum of collective excitations and power-law correlations with $K$-dependent critical exponents. As follows from the definitions Eqs. \eqref{tot-den}-\eqref{D1-2-bos}, strong quantum fluctuations suppress any kind of ordering in the system, including charge imbalance between the legs of the ladder,  dimerization along the zigzag links, and net current in the ground state. Within the range $1/2 <K<1$, the $\tau$-proportional part  of the relative density Eq.~\eqref{rel-den} contributes to dominant correlations in the model: at 
 distances 
$
|x| > \alpha \left( K |\tau| \right)^{-\frac{1}{2(1-K)}}
$
the corresponding correlation function follows the power law 
\begin{equation}
\label{TK-rr-corr}
    \langle  \rho_{\text{rel}} (x)\rho_\text{rel} (0)\rangle
        \simeq \frac{1}{2(\pi \alpha)^{2}}
            \frac{\tau^{2}}{1+\tau^{2}}
            \left( \frac{\alpha}{|x|} \right)^{2K}.
\end{equation}

At $K<1/2$, the $\lambda$-perturbation in Eq.~\eqref{eq:SG11} becomes relevant and the model flows towards strong-coupling with a dynamical generation of a mass gap

\begin{equation}
\label{eq:mass-gap-lambda}
    m_{\lambda} \sim
        \frac{v_{c}}{\alpha}
        \left(\frac{|\lambda|}{v_{c}}\right)^{1/2(1-2K)},
    \quad
    (K<1/2).
\end{equation}
Since $\lambda > 0$, the field $\Phi$ is locked in one of the infinitely degenerate vacuum values
\begin{equation}
\label{lambda-vacua1}
    \left( \Phi  \right)_{\text{l}} =
        \frac{1}{2}\sqrt{\frac{\pi}{K}}l,
    \quad
    l\in\mathbb{Z}
\end{equation}
Therefore $\langle B_{1} \rangle = \langle B_{2} \rangle = 0$, but the average relative density turns out to be nonzero:
\begin{equation}
\label{ro-rel-mu01}
    \langle \rho_{\text{rel}} \rangle_{\lambda} = \pm \rho_{0},
    \quad
    \rho_0 \sim |\tau|^{\frac{1-K}{1-2K}}.
\end{equation}
According to Eqs.~\eqref{rel-den}-\eqref{j1-j2-bos}, the population imbalance between the chains is accompanied by a spontaneous generation of a net current 
\begin{eqnarray}
    \langle j_0 \rangle_{\lambda} =
        \langle j_z \rangle_{\lambda} =
        \frac{v_{\text{F}}}{\pi \alpha}\frac{\tau}{1+\tau^{2}}
        \langle :\cos \sqrt{4\pi K}\Phi(x): \rangle_{\lambda}.
\end{eqnarray}
The spontaneous relative density, and  hence the current, are non-analytic functions of the frustration parameter $\tau$. Being zero at $K>1/2$, they  exponentially increase on decreasing $K$ in the region $K < 1/2$ following the law Eq.~\eqref{ro-rel-mu01}. The quantum phase transition taking place at $K=1/2$ belongs to the Berezinskii-Kosterlitz-Thouless (BKT) universality class \cite{Nersesyan2004,Giamarchi2003}.

Elementary excitations of the charge-transfer phase are topological quantum solitons of the SG model Eq.~\eqref{eq:SG11}. They carry the mass given by Eq.~\eqref{eq:mass-gap-lambda} and  a fractional  fermionic number $Q_s=1/2$. This number is identified with the topological charge of the soliton which interpolates between neighboring vacua of the cosine potential $\cos \sqrt{16\pi K} \Phi$:
\begin{equation}
    Q_s = \sqrt{\frac{K}{\pi}} 
        \int_{-\infty}^{\infty} \mathrm{d} x\ \partial_{x} \Phi(x)
        = \frac{1}{2}.
        \label{frac-number}
\end{equation}

\subsubsection{$m \neq 0$ regime with $K>\frac{1}{2}$}
\label{subsection:band-insulator}

At $K>1/2$ and $m \neq 0$ the interband pair-hopping processes are irrelevant, and 
the effective theory is given by the SG model:
\begin{equation}
\label{eq:m-SG-A}
    \begin{split}
        \mathcal{H} (x) =&
            \frac{v_{c}}{2} \left[ \Pi^{2} (x) +
            \left( \partial_{x} \Phi (x) \right)^{2} \right]
        \\
        &+ \left(\frac{m}{\pi \alpha}\right)
        :\sin \sqrt{4\pi K} \Phi (x) : .
    \end{split}
\end{equation}
It describes a bosonized version of the theory of marginally perturbed
massive fermions -- the so-called massive Thirring model~\cite{Coleman1975}. The scalar field $\Phi$ is locked in one of the infinitely degenerate vacua
\begin{equation}
\label{eq:fi-vacua-A}
    \left( \Phi  \right)^{\rm vac}_{l} =
        \sqrt{\frac{\pi}{K}}
        \left( - \frac{1}{4} \text{sgn}(m) + l \right) ,
    \quad
    l\in\mathbb{Z}.
\end{equation}
Eq.~\eqref{eq:fi-vacua-A} displays two subsets of the vacua corresponding to different signs of $m$, each subset describing a band insulator. The excitation spectrum has a mass gap ${m}_s$:
\begin{equation}
\label{gen-mass}
    m_{s} = C(K) \left(\frac{v_{c}}{\alpha} \right) 
        \left(  \frac{|m|\alpha}{v_{c}} \right)^{1/(2-K)}
        \mathrm{sgn}(m),
\end{equation}
where $C(K)$ is a dimensionless constant tending to $1$ as $K \to 1$.
The mass term in Eq.~\eqref{eq:m-SG-A} explicitly breaks parity and, according to Eq.~\eqref{eq:D-rot} and Eq.~\eqref{D1-2-bos}, leads to 
a finite dimerization $\langle B \rangle $ of the zigzag bonds of the ladder:
\begin{equation}
\label{D-ave}
    \langle B \rangle 
        \sim |m_{s}|^K \mathrm{sgn}(m) 
        \sim |m| ^{K/(2-K)} \mathrm{sgn}(m) .
\end{equation}

The quantum soliton of the SG model Eq.~\eqref{eq:m-SG-A} carries the mass $m_{s}$ and topological charge $Q_{\text{F}} = 1$ and thus is identified as the fundamental fermion of the related massive Thirring model~\cite{Coleman1975}. In the ground state $\langle \partial_{x} \Phi \rangle = \langle \partial_{x} \Theta \rangle = 0$. Moreover, for the vacuum values of the field $\Phi$ given by Eq.~\eqref{eq:fi-vacua-A} the average $\langle \cos \sqrt{4\pi K} \Phi \rangle$ vanishes. So at $\rho=1$ and $m\neq 0$ the total and relative densities remain unaffected by the flux. Correlations of the relative density are short-ranged.

Thus, at $m\neq 0$ the SG model Eq.~\eqref{eq:m-SG-A} describes band insulator phases. Their thermodynamic properties depend only on the magnitude of the spectral gap $|m_{s}|$, while their topological properties are determined by the sign of the bare mass $m$. With the sign of the ``heavy'' mass fixed ($M>0$), the band insulator phase at $m>0$ is topologically trivial, whereas the corresponding phase at $m<0$ is topologically nontrivial. This conclusion has been reached in Ref.~\cite{Beradze2023} by inspecting the 2$\times$2 matrix structure of the Bloch Hamiltonian describing bulk properties of the non-interacting model. In the present paper, we provide extra support to this conclusion by studying boundary zero-energy states of a semi-infinite triangular asymmetric flux ladder. The corresponding calculations are given in Appendix \ref{appendix:boundary}.

A more complete characterization of the two band insulating phases described by the SG model in Eq.~\eqref{eq:m-SG-A} is extracted from the mass dependence of non-local order parameters~\cite{Hida1992} -- the parity $(\hat{O}_{P}) $ and string-order $(\hat{O}_{S})$  operators:
\begin{eqnarray}
    \hat{O}_{P} (j) = \exp( i \pi \sum_{k \leq j} \delta\hat{n}_k), ~~~~\hat{O}_{S} (j) = \hat{O}_{P} (j)  \delta \hat{n}_j. 
    \label{O-PS}
\end{eqnarray}
Here $j$ labels the diatomic unit cells of the ladder, and  $\delta \hat{n}_j = \sum_{\sigma} :c^{\dagger}_{j\sigma} c_{j\sigma}:$ is the density fluctuation on the zigzag rung $j$. Non-local string and parity orders have been considered earlier for interacting bosons~\cite{dalla-2006, Berg2008, Batrouni2013, Batrouni2014} to specify the differences between the Mott and Haldane insulator gapped phases. It was shown that the two non-local order parameters are dual to each other \cite{Berg2008}: $\langle \hat{O}_{P}\rangle \neq 0$, $\langle \hat{O}_{S} \rangle = 0$ for the  Mott insulator, and $\langle \hat{O}_{P}\rangle = 0$, $\langle \hat{O}_{S} \rangle \neq 0$ for the Haldane insulator. Non-local parity and string order have been shown to characterize strongly correlated states of interacting fermions in strictly one-dimensional systems \cite{montorsi-12, montorsi-13}, as well as in fermionic ladder models \cite{Bahri2014, Chitov2017, ners-2020}. A significant progress in the direct measurement of non-local parity and string correlations in low-dimensional ultra-cold Fermi and Bose systems has recently been reported~\cite{Endres2011, Hilker, Lesleuc2019, Sompet2022, Wei2023}.

The string and parity order parameters, Eqs.~\eqref{O-PS}, being non-local in terms of the densities $\hat{n}_k$, admit a local representation in terms of vertex operators of the effective SG model. In both Bose  \cite{Berg2008} and Fermi  \cite {ners-2020}  cases, the perturbation to the Gaussian Hamiltonian for the field $\Phi$ was defined as $m \cos{\sqrt{4\pi K} \Phi}$, with the Luttinger-liquid constant $K$ being close to 1. Using Abelian bosonization, with such a definition, one shows that in the continuum limit
\begin{eqnarray}
\label{nonlocal-bos-conven}
    \hat{O}_{P} (x) \sim \sin \sqrt{\pi K} \Phi(x), ~~~ \hat{O}_{S} (x) \sim \cos \sqrt{\pi K} \Phi(x).
\end{eqnarray}
However, in the SG model of Eq.~\eqref{eq:m-SG-A} the perturbation is of the form $m \sin{\sqrt{4\pi K} \Phi}$. Therefore the bosonic representation of the operators $\hat{O}_{P}$ and $\hat{O}_{S}$ must be modified. The sine transforms to a cosine under the shift $\Phi \to \Phi - (1/4) \sqrt{\pi/K}$, implying the
following the redefinition of the non-local order parameters
\begin{equation}
\label{new-OPS}
    \begin{split}
        \hat{O}_{P} &\to \frac{1}{\sqrt{2}} 
            \left(
                \sin \sqrt{\pi K} \Phi - \cos{\sqrt{\pi K} \Phi}
            \right),\\
        \hat{O}_{S} &\to \frac{1}{\sqrt{2}} 
            \left(
                \cos \sqrt{\pi K} \Phi + \sin{\sqrt{\pi K} \Phi} 
            \right).
    \end{split}
\end{equation}
Using the fact that in the ground state of the SG model~\eqref{eq:m-SG-A} the vacuum values of the field $\Phi$ are given by Eq.~\eqref{eq:fi-vacua-A}, we conclude that the expectation values of $\hat{O}_{P}$ and $\hat{O}_{S}$ are given by
\begin{eqnarray}
    \begin{split}
        \langle \hat{O}_{P} (m; l) \rangle &=  (-1)^{l-1}  F(m)\theta (m),\\
        \langle \hat{O}_{S} (m; l) \rangle &= (-1)^l F(m) \theta (-m),
    \end{split}
    \quad
    l \in \mathbb{Z},
\end{eqnarray}
where at $|m|\alpha/v \ll 1$ ( i.e., in the vicinity of the Gaussian line $m=0$) and
\[
F(m) \sim \left(\frac{|m|\alpha}{v} \right) ^{\frac{K}{4(2-K)}}.
\]
Thus, the band insulator phase with $m>0 ~(t_1 > t_2)$ is topologically trivial ($\langle \hat{O}_{S}\rangle = 0$) and characterized by a non-zero non-local parity order $(\langle \hat{O}_{P} \rangle \neq 0)$, whereas the phase with $m<0 ~(t_1 < t_2)$ represents a topological insulator with a non-zero string order ($\langle \hat{O}_{S}\rangle \neq 0$) but lacks parity order ($\langle \hat{O}_{P}\rangle = 0$). Below, in Sec.~\ref{sec:num_phases}, we provide numerical evidence for this conclusion.

\subsubsection{$m \neq 0$ regime with $K<\frac{1}{2}$}
\label{sec:subsubsec3}

\begin{figure}[h]
    \centering
        \includegraphics[width=3in]{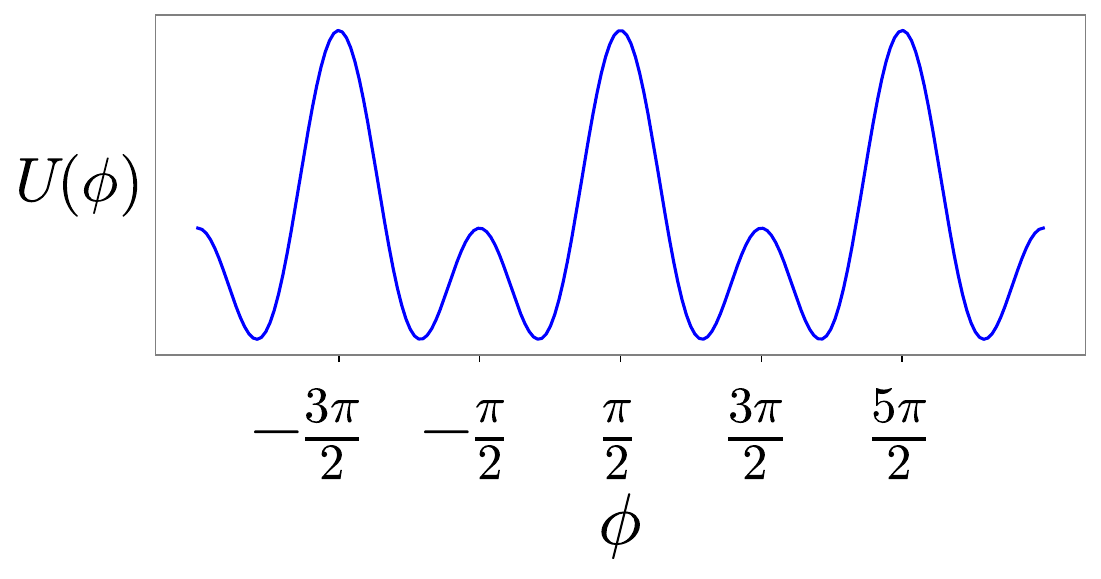}
    \caption{
        A pictorial representation of the potential  $U(\phi) = a \sin \phi - \cos 2\phi$,
        $\phi = \sqrt{4\pi K} \Phi$,
        $a=(2\pi \alpha)m/\lambda$.
    } 
    \label{fig:2well}
\end{figure}

The most interesting situation arises when  both perturbations of the DSG model in Eq.~\eqref{eq:H-dsg} are  relevant. This case can  be realized when the interchain repulsion satisfies the condition $g \geq g_c$, where $g_c$ is a non-universal value of the coupling constant such that $K(g_c) = 1/2$. At $\lambda > 0$ the DSG potential  of Eq.~\eqref{eq:H-dsg} is shown in Fig.~\ref{fig:2well}. It represents a sequence of double-well potentials of the $\mathbb{Z}_2$-symmetric Ginzburg-Landau theory  with the minima merging into $\phi^4$ local wells at $2\pi \alpha m /\lambda = 4$. These semi-classical considerations lead to  the conclusion that at $\lambda >0$ the interplay of the two perturbations of the DSG model is resolved as the appearance of quantum criticalities belonging to the Ising universality class with central charge $c=1/2$~\cite{Delfino1998}. 
In the following Sec.~\ref{sec:numerics}, we confirm the existence of $c=1/2$ criticalities for $m \neq 0$ at critical interaction strengths $V=V_c(m)$.

As pointed out in Ref.~\cite{Delfino1998}, at a quantum level, the infrared behavior of the DSG model is determined by the ratio of the mass gaps separately generated by each of the two perturbations in the absence of the other. The parameter $\varrho \sim|{m_{s}}/m_{\lambda}|$ controls the two perturbative regimes of the DSG model (\ref{eq:H-dsg}): $\varrho \to0$ and $\varrho \to \infty$. The Ising transitions occur in a non-perturbative region where the two masses are of the same order. The condition $\varrho \sim 1$ gives an order-of-magnitude estimate of the critical curves on the plane ($\lambda, m$):
\begin{equation}
m = \pm m^* (\lambda), ~~~m^* (\lambda) \sim \frac{v_{c}}{\alpha} \left( \frac{\lambda}{v_{c}}  \right)^{\frac{2-K}{2(1-2K)}}.
\label{crit-curve}
\end{equation}

The two critical lines Eq.~\eqref{crit-curve} separate the band-insulator phases, $|m| > m^* (\lambda)$, from
the mixed phase occupying the region $-m^* (\lambda) < m < m^* (\lambda)$,
in which charge imbalance coexists with dimerization of the zigzag bonds.  As shown in Fig.~\ref{fig:2well}, in the mixed  phase the minima of the DSG potential decouple into two subsets
\begin{equation}
\label{eq:dw-minima}
    \begin{split}
        \left( \Phi \right)^{\text{odd}}_{k}
            &= \frac{1}{2} \sqrt{\frac{\pi}{K}}
            \left( 2k+1 + \frac{\eta}{\pi} \right)
        \\
        \left( \Phi \right)^{\text{eve}n}_{k}
            &= \frac{1}{2} \sqrt{\frac{\pi}{K}}
            \left( 2k -\frac{\eta}{\pi} \right),
    \end{split}
\end{equation}
where $k=0, \pm 1, \pm 2, ...$ and $\eta = a/4 = 2\pi\alpha m/4\lambda$. Accordingly, there are two types of massive topological excitations carrying $\eta$-dependent fractional charge. These kinks are
associated with the vacuum-vacuum interset transitions
$
\left( \Phi \right)^{\text{odd}}_{k} ~\leftrightarrow~\left( \Phi \right)^{\text{even}}_{k, k+1}.
$
The long kinks carry the charge $Q_+ = {1}/{2} + {|\eta|}/{\pi}$ and interpolate between the solitons of the $m=0$ phase with spontaneously broken $\mathcal{P}$-symmetry ($Q=1/2$) and single-fermion excitations of the band-insulator phase ($Q=1$). On the other hand, on approaching the Ising criticality, the short kinks with the topological quantum number $Q_- = {1}/{2} - {|\eta|}/{\pi}$ lose their charge and mass and transform to a neutral collective excitonic mode.

According to the analysis done in \cite{Fabrizio2000}, in the vicinity of the critical lines Eq.~\eqref{crit-curve} the gapped phases in the regions $|m|> m^*$ and $|m| < m^*$ are identified as Ising ordered and disordered phases, respectively. Near the transition the average relative density $:\rho_\text{rel} (x):$ behaves as the  disorder operator $\mu(x)$ of the underlying quantum Ising model \cite{Fabrizio2000}. Therefore, the phases realized at $|m| > m^*$ are the already discussed zigzag-dimerized insulating phases with equally populated chains: $\langle :\rho_\text{rel}: \rangle= 0$. At $|m| < |m^*|$ the relative density acquires a finite expectation value which vanishes as
\begin{equation}
    \langle  :\rho_{\text{rel}}: \rangle \sim (|m^*| - |m|)^{1/8}
    \label{ro-ising}
\end{equation}
on approaching the critical point from below: $|m| \to |m^*| - 0$. 

The two Ising critical lines merge at the point $m=0$, $K=1/2$ and, as shown in Fig.~\ref{fig:Ashkin-Teller-Phase-Diagram} transform to a Gaussian critical line $m=0$, $K>1/2$ which describes a Tomonaga-Luttinger liquid phase. In the next section, we show that the merging point of the Ising critical lines represents a bifurcation point of the Ashkin-Teller (AT) model where the symmetry is enlarged to SU(2).

\section{Ashkin-Teller and double-frequency sine-Gordon models: Equivalence to a weakly dimerized XXZ spin-1/2 chain}
\label{sec:Ashkin-Teller}

In this section, we elaborate  on the relationship between the quantum 1D version of the AT model, staggered XXZ spin-1/2 chain, and the DSG model of Eq.~\eqref{eq:H-dsg} at $K$ close to 1/2. The relation between the AT model, considered in the scaling limit, and the DSG model of a scalar field has been anticipated in  earlier studies \cite{Delfino1998, Fabrizio2000, Kadanoff1981, Delfino2004}. Here we focus on the connection between the above two models on one hand, and an effective isotropic S=1/2 Heisenberg chain, weakly perturbed by an explicit dimerization and exchange anisotropy, on the other.

The classical  AT model describes two identical 2D Ising models coupled by a four-spin interaction. 
As is well known~\cite{Kogut1979}, by virtue of transfer matrix formalism, two-dimensional classical statistics can be viewed as an imaginary-time (i.e., Euclidean) version of quantum mechanics in 1+1 dimensions. 
The quantum lattice version of the AT model was derived in~\cite{Kohmoto1981}: 

\begin{equation}
\label{eq:quantumAT}
    \begin{split}
        H_{\rm QAT}=- \frac{1}{4}\sum_{j=1}^N
        \Big [
            &\left(
                J_{+}\sigma^{z}_{1,j} \sigma^{z} _{1,j+1} +
                J_{-} \sigma^{x}_{1,j}
            \right) + 
            \\
            &\left(
                J_{+} \sigma^{z}_{2,j} \sigma^{z} _{2,j+1} +
                J_{-} \sigma^{x}_{2,j}
            \right)+
            \\
            &q
            \left(
                J_{+}
                \sigma^{z} _{1,j}\sigma^{z}_{1,j+1}
                \sigma^{z} _{2,j}\sigma^{z}_{2,j+1}
                + J_{-}\sigma^{x}_{1,j} \sigma^{x}_{2,j}
            \right)
        \Big].
    \end{split}
\end{equation}
The Hamiltonian \eqref{eq:quantumAT} describes two coupled quantum Ising chains. 
The relationship between the constants $J_{\pm}$, $q$ and main parameters of the 2D AT model can be found
in ~\cite{Kohmoto1981}.
At $q = 1$ $H_{\rm QAT}$
possesses a hidden SU(2) symmetry. 
Indeed, using a specially designed nonlocal unitary transformation it has been shown \cite{Kohmoto1981}
that the quantum AT model (\ref{eq:quantumAT}) is  exactly equivalent to a model of a staggered XXZ spin-1/2 chain:
\begin{equation}
\label{eq:stag-xxz}
    H_{\rm S} = \sum_{n=1}^{2N}
        \left[ J_{0} + (-1)^n J_{1} \right]
        \left(
            \textbf{S}_{n}\cdot \textbf{S}_{n+1} +
            \rho S^{z}_{n} S^{z}_{n+1}
        \right),
\end{equation}
where $J_{0,1} = (J_+ \pm J_-)/2$, $\rho = q-1$.

At $|J_{1}|,|\rho| \ll J$, the  model in Eq.~\eqref{eq:stag-xxz} occurs in the vicinity of the isotropic Heisenberg point, $J_{1}= \rho =0$, where it is critical with the central charge $c=1$ and whose properties in the scaling limit are described by the critical SU(2)$_1$ WZNW model with a marginally irrelevant perturbation \cite{Affleck1986}:
\begin{equation}
\label{eq:wznw+}
    \mathcal{H}_{0}  =
    \frac{2\pi v_{s}}{3} \left( :\textbf{J}^2 _R: + :\textbf{J}^2 _L:  \right)
     - g_0 \textbf{J}_R \cdot \textbf{J}_L,
\end{equation}
with $g_0 \sim v_{s} \sim J_0 a_0 > 0$. Here $\textbf{J}_{\text{R,L}}$ are the generators of the chiral,
level-1 SU(2) Kac-Moody algebra (see for details the textbooks \cite{DiFrancesco1997,Mussardo2010}).
A finite $\rho$-term in (\ref{eq:stag-xxz}) introduces exchange anisotropy. 
The translationally invariant chain ($J_{1} = 0$) with
 $\rho < 0$ occurs in a Tomonaga-Luttinger liquid phase with $\rho$-dependent critical exponents \cite{Luther1975}, whereas at $\rho > 0$ the system enters a gapped Neel phase with a doubly degenerate ground state \cite{Haldane1982}. The Neel ordering is site-parity ($P_{S}$) symmetric but breaks spontaneously link parity ($P_{L}$). At $\rho = 0$, $J_{1} \neq 0$ the chain maintains spin-rotational  symmetry but is explicitly dimerized. Its spectrum is massive. The $J_{1}$-perturbation breaks $P_{S}$ but preserves $P_{L}$.

So, there are two, mutually incompatible by symmetry, ``massive'' directions at the SU(2) critical point parametrized by the couplings $J_1$ and $\rho$. Their competition gives rise to the splitting of the SU(2)$_1$ WZNW criticality into two Ising criticalities. For small values of $J_1$ and $\rho$ the low-energy properties of the model in Eq.~\eqref{eq:stag-xxz} with both perturbations present can be adequately described in terms of a perturbed Gaussian
theory with the structure of the DSG model \cite{Kohmoto1981}:
\begin{equation}
\label{eq:DSG-AT}
    \begin{split}
        \mathcal{H}_{\text{DSG}}
            & = \frac{u_s}{2} \left[ \left( \partial_x \Phi\right)^2 + \left( \partial_x \Theta\right)^2\right]
            \\
            & +\frac{h }{\pi\alpha} \sin \sqrt{2\pi K_{s}} \Phi
            - \frac{g_{\perp}}{(2\pi \alpha)^2} \cos \sqrt{8\pi K_{s}} \Phi,
    \end{split}
\end{equation}
This mapping is valid up to irrelevant corrections.
In Eq. (\ref{eq:DSG-AT})  $u_s$ is a renormalized velocity, $h \sim J_{1}$,
\begin{equation}
\label{eq:Ks}
    K_{s} = 
        \left(
            \frac{1 - g_{\parallel}/4\pi v_{s}}
            {1 + g_{\parallel}/4\pi v_{s}}
        \right)^{1/2}
        \simeq 1- g_{\parallel}/4\pi v_{s} + \dots, 
\end{equation}
where $g_{\parallel} = g_{0} - C_{1} \alpha \rho$, $C_{1}>0$ being a nonuniversal numerical constant; the coupling constant
$g_{\perp} = g_{0} + C_{2} \alpha \rho$, $C_{2} > 0$ being another constant.

We observe that, even though there
is no direct mapping of the spin models (\ref{eq:stag-xxz}) or (\ref{eq:quantumAT})
onto the original fermionic model (\ref{eq:FullHamiltonian}), or vice versa,
under the identifications
\begin{equation}
    h = m, \quad g_{\perp} = 2\lambda, \quad K_{s} = 2K \label{2sg-param}
\end{equation}
the DSG models in Eq.~\eqref{eq:DSG-AT} and Eq.~\eqref{eq:H-dsg} coincide. Thus we conclude that
at
$K\sim 1/2$ and small $m$, the DSG model \eqref{eq:H-dsg}, being derived as 
a field-theoretical limit of the original flux-ladder model, at the same time describes scaling properties
a weakly dimerized spin-1/2 chain with a small exchange anisotropy.
Let us stress again, that while the QAT and spin-chain lattice Hamiltonians, Eqs.(\ref{eq:quantumAT}) and ({\ref{eq:stag-xxz}}),
are unitarily equivalent, the correspondence between the spin-chain Hamiltonian $H_{\rm S}$ and the field-theoretical model
$H_{\rm DSG}$ in (\ref{eq:DSG-AT}) only holds in the scaling limit. This fact renders the SU(2) symmetry of our fermionic ladder model
at the bifurcation point an \textit{emergent} phenomenon.

Although the above discussion of model Eq.~\eqref{eq:DSG-AT} 
concerned the vicinity of the XXX point, this model maintains its applicability to a broader region of the parameter space where $K_{s} < 1$. In particular, there exists a `decoupling' point $K_{s} = 1/2$ where the DSG model
can be mapped onto two non-critical (disordered) Ising models coupled by an interaction $h \sigma_1 \sigma_2$ \cite{Delfino1998,Fabrizio2000}. However, in that region, the two mutually dual Ising critical lines are well separated and their merging point is not accessible. On the contrary, the present discussion, relying on the equivalence with the staggered XXZ spin chain, treats the DSG model as a weakly perturbed SU(2)$_1$ WZNW model. In such an approach, the SU(2) symmetry emerging at the bifurcation point $K_{s} = 2K = 1$ of Gaussian line $m=0$ into two Ising critical lines finds its natural explanation. 

Thus, as follows from the above discussion, mapping of the original spinless fermionic flux-ladder model to the DSG field theory plays a central role in the present paper.
While we are not in a position to determine the (non-local) generators of the hidden symmetry emerging at the critical point, the field theory analogy is fully consistent with all other our field theory predictions: we will then proceed in the next section with a numerical verification of our findings. We would like to note that our case differs substantially from cases with emergent continuous Abelian symmetries, that are already relatively well understood (for some recent examples, see Ref.~\cite{peotta2014xyz,giudici2019diagnosing,Jouini2023}). 


\section{Numerical treatment}
\label{sec:numerics}

\begin{figure}[tb]
    \includegraphics[width=\linewidth]{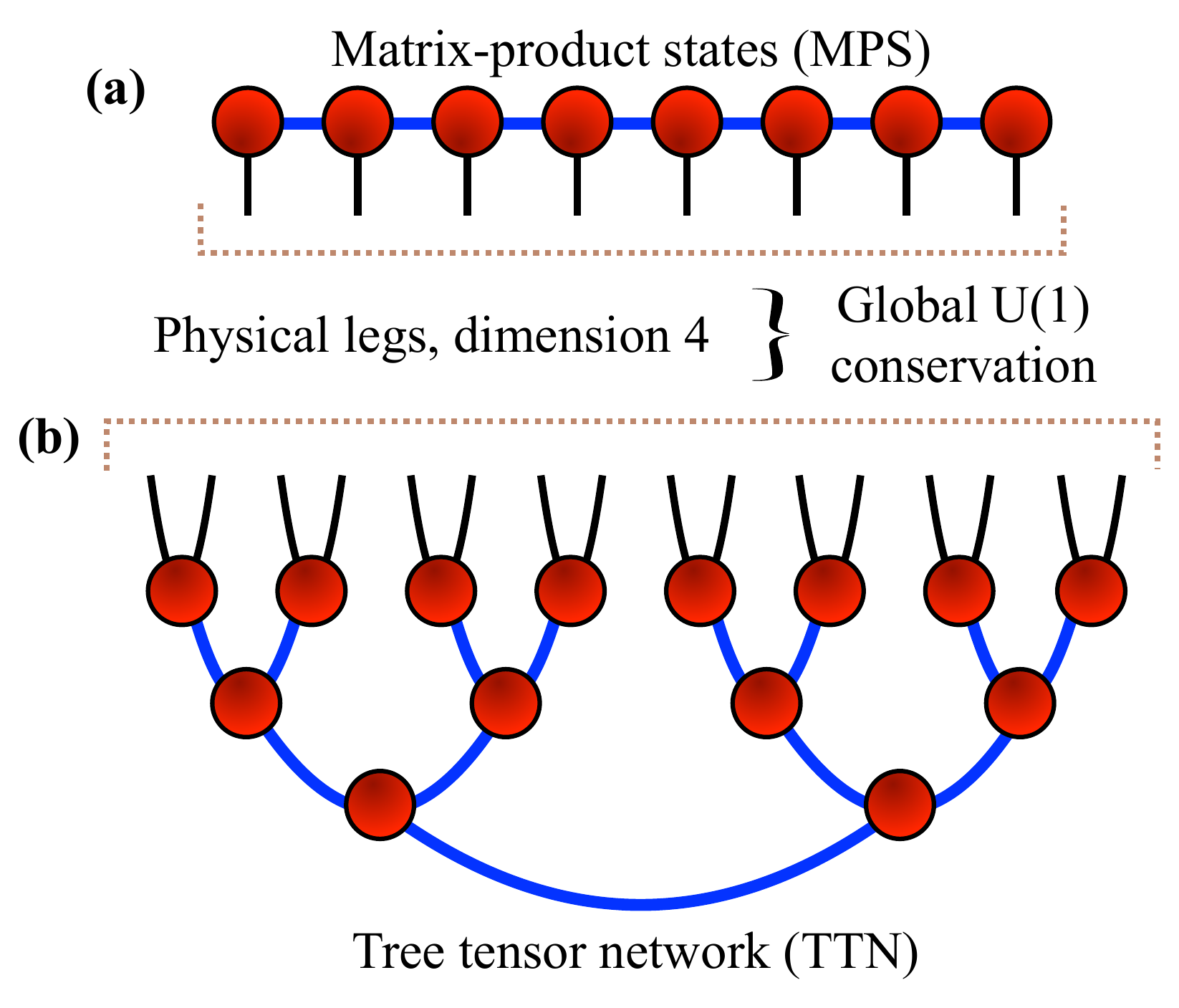}
    \caption{(Color online.) Tensor network (TN) ansatzes used in our numerical simulations.  We use two different TN ansatzes, namely (a) the matrix-product state (MPS) and (b) the tree tensor network (TTN), for our analysis. In each case, the physical dimension is four, and we employ $\text{U(1)}$ symmetric tensors~\cite{Singh_PRA_2010,Singh_PRB_2011} to conserve the total particle number.
    In our simulations, we group two sites along the rungs of the ladder (following the labeling in Fig.~\ref{fig:zigzag}) to define the physical sites of the TN states.
    }
    \label{fig:tn_ansatz}
\end{figure}

To validate the analytical approaches and extend the prediction to larger coupling strengths, we now employ state-of-the-art tensor-network (TN) simulations, see Fig.~\ref{fig:tn_ansatz}.
To mitigate any finite-size boundary effects, we evaluate the system either at the thermodynamic limit, or at finite sizes with periodic boundary conditions (PBC), unless stated otherwise.

For infinite lattices, we employ the infinite density-matrix renormalization group (iDMRG) technique~\cite{white_prl_1992, white_prb_1993, white_prb_2005, McCulloch_2007, McCulloch_2008, Crosswhite_2008} based on the matrix-product state (MPS) ansatz~\cite{schollwock_aop_2011, Orus_aop_2014} (Fig.~\ref{fig:tn_ansatz}(a)). Specifically, we use the infinite variation of MPS known as the iMPS~\cite{Vidal_2007, Kjall_PRB_2013}. 
For the finite system-sizes with PBC, we apply tree tensor network (TTN) methods~\cite{Tagliacozzo2009, Gerster2014,Silvi2019} (Fig.~\ref{fig:tn_ansatz}(b)), which can, unlike MPS, handle PBC with similar computational cost and accuracy as open boundary conditions (OBC)~\cite{Gerster2014}. 
In the following, unless otherwise stated, we fix $t_0=1$ to set the energy unit of the system, and consider $M = t_{1}+t_{2} = 0.2 < t_0$.
We also consider the situation of repulsive interchain interaction, i.e., $V \geq 0$,
and we analyze the phase diagram in the $(m/t_0, V/t_0)$ parameter space. The regime of repulsive interaction corresponds to $\lambda >0$, via Eq.~\eqref{eq:couplings}. For the scenario of attractive interactions,  see Appendix~\ref{appendix:attrac}.

\begin{figure*}[htb]
    \includegraphics[width=\linewidth]{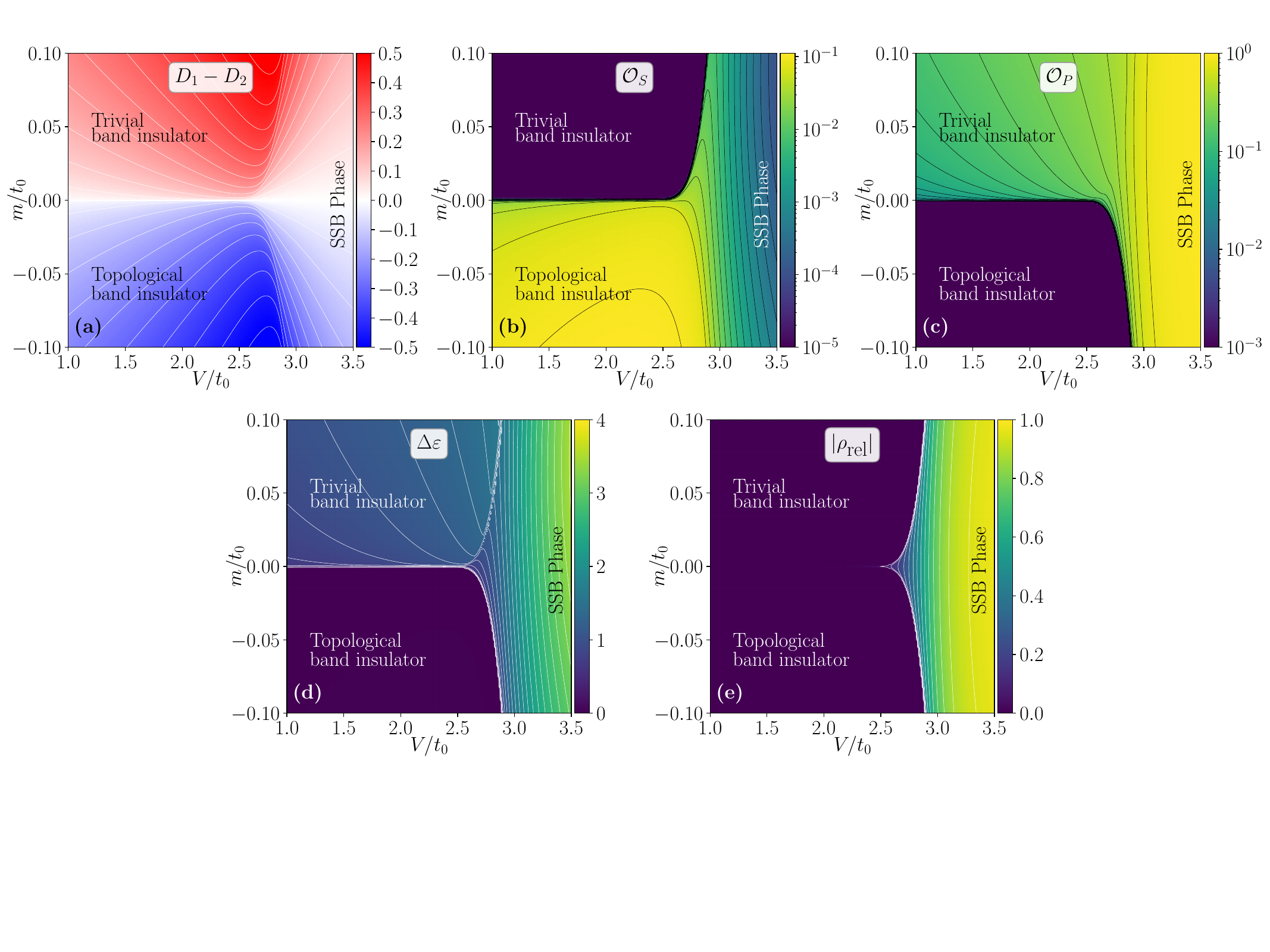}
    \caption{(Color online.) The characterization of different phases of the repulsive triangular ladder in the $(m/t_0, V/t_0 >0)$-plane.
    (a) We consider the difference between two different rung tunneling amplitudes $D_1 = \overline{\braket{c_{j,+}^{\dag} c_{j,-} + \text{h.c.}}}$ 
    and $D_2 = \overline{\braket{c_{j,+}^{\dag} c_{j-1,-} + \text{h.c.}}}$ respectively. This difference highlights the different types of zigzag dimerization in the band insulator phases.
    (b)-(c) The non-local string ($\mathcal{O}_S$) and parity ($\mathcal{O}_P$) correlation functions as defined in Eqs.~\eqref{eq:string_parity}. $\mathcal{O}_S$ is zero (non-zero) while $\mathcal{O}_P$ is non-zero (zero) in the trivial (topological) band insulator phase. Both become non-vanishing in the large-$V/t_0$ symmetry-broken phase.
    (d) The entanglement gap $\Delta \varepsilon = \varepsilon_1 - \varepsilon_0$, where $\varepsilon_0$ and $\varepsilon_1$ are the ground and first excited state energies of the entanglement Hamiltonian $H^{E}_l = -\ln \rho_l$ respectively,  is plotted in the parameter space. Vanishing values $\Delta \varepsilon$ in the band insulator phase for $m < 0$ dictates the topological nature of this phase.
    (e) The relative density $\rho_{\text{rel}} = \overline{\braket{\hat n_{j, +} - \hat n_{j, -}}}$ between the legs serves as a order parameter for the $\mathbb{Z}_2$ symmetry-breaking associated with $\mathcal{P}$ symmetry. The order parameter $\rho_{\text{rel}}$ becomes non-vanishing in the two-fold degenerate $\mathbb{Z}_2$-broken phase for $V > V_{c}$.
    Here, we have used iDMRG simulations with bond dimension $\chi=256$.
    }
    \label{fig:idmrg_orders}
\end{figure*}

\subsection{Phase diagram}
\label{sec:num_phase_diagram}

To determine different phase transitions and differentiate different phases, we perform our numerical simulations over the $(m/t_0, V/t_0)$-plane and first consider
the system correlation length $\xi$.
The correlation length $\xi_O$ corresponding
to any local operator $O_j$ is defined by the length scale associated with the correlation function: 
\begin{equation}
    \langle O_j O_{j+R} \rangle - \braket{O_j} \braket{O_{j+R}} \sim \exp(-R/\xi_O).
\end{equation}
Then the system correlation length $\xi$ of the quantum state is given by the maximum of these length scales as 
\begin{equation}
    \xi=\max(\xi_{O_1},\xi_{O_2},\cdots).
\end{equation}
For an iMPS ground state with bond dimension $\chi$, the correlation length is $\xi_{\chi} = -1/\ln|\epsilon_2|$, where $\epsilon_2$ is the second largest eigenvalue of the iMPS transfer matrix~\cite{Kjall_PRB_2013}. It is to be noted that in case of critical systems where the system correlation length diverges, $\xi_{\chi}$ is the length-scale artificially introduced by the finite iMPS bond dimension $\chi$ and usually $\xi_{\chi} \sim \chi^{\beta}$, with $\beta$ being a scaling exponent.

In Fig.~\ref{fig:Ashkin-Teller-Phase-Diagram}(b), we show the phase-diagram of the system in the $(m/t_0, V/t_0)$-plane through the lens of correlation length for iMPS bond dimension $\chi=256$. Clearly, we see a bifurcation of critical line at $m=0$ into two critical lines at around $V/t_0 \simeq 2.45$ similar to what has been seen in the Ashkin-Teller (AT) model, see Sec.~\ref{sec:subsubsec3} and Sec.~\ref{sec:Ashkin-Teller}. Although the bosonization approach is controlled only for weak coupling regime $V/t_0 \ll 1$, it predicted, from the phenomenological treatment of the double-frequency sine-Gordon (DSG) model, the existence of the bifurcation point (Sec.~\ref{sec:Ashkin-Teller}), that appeared at relatively strong coupling regime $V/t_0 \simeq 2.45$. 
Below, we show that this bifurcation is indeed of the $\text{SU(2)}_1$ WZNW type, where a Gaussian  critical line with central charge $c=1$ (at $m=0$ and $V/t_0 \lesssim 2.45$) bifurcates into two Ising transitions with $c=1/2$ (for $m \neq 0$ and $V/t_0 \gtrsim 2.45$).
Moreover, Fig.~\ref{fig:Ashkin-Teller-Phase-Diagram}(b) also suggest that apart from the critical lines, all the three phases are gapped as they possess finite correlation lengths.

\begin{figure}[tb]
    \includegraphics[width=\linewidth]{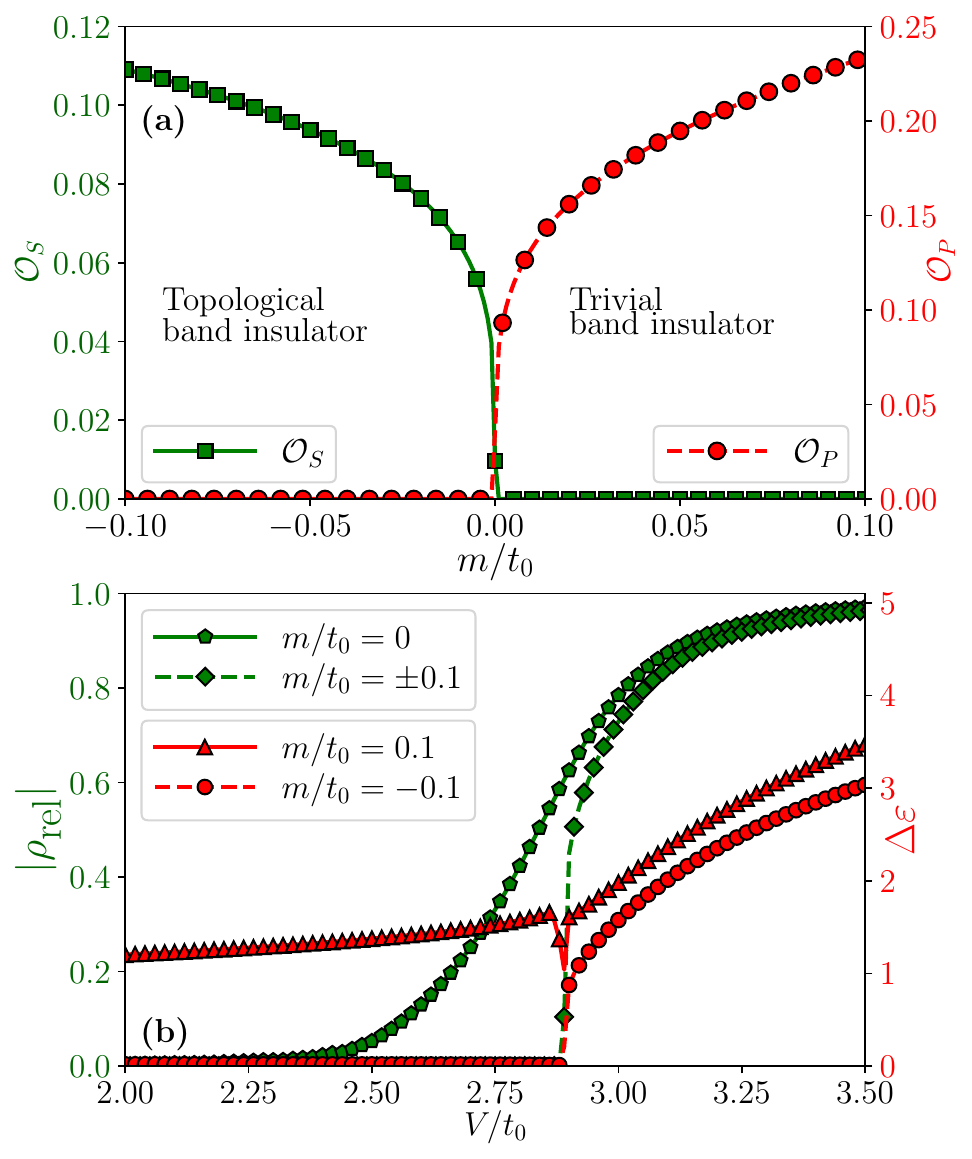}
    \caption{(Color online.) 
    (a) The string $\mathcal{O}_S$ and parity $\mathcal{O}_P$ correlations for varying $m/t_0$ and fixed $V/t_0=2$.
    (b) The variations of the order parameter $\rho_{\text{rel}}$ and the entanglement gap $\Delta \varepsilon$ with varying $V/t_0$ for fixed values of $m/t_0$ as indicated in the figure. All other details are the same as in the Fig.~\ref{fig:idmrg_orders}.
    }
    \label{fig:idmrg_orders_cut}
\end{figure}

\subsection{Characterization of different phases}
\label{sec:num_phases}

The band insulator phases at lower values of $V/t_0$ undergo distinct types of zigzag dimerization as explained in Sec.~\ref{subsection:band-insulator}. For $m>0$, the dimerization occurs along $t_{1}$ links and the phase is a trivial band insulator. For $m<0$ regime, the dimerization is along $t_{2}$ links, where the phase is a topological band insulator. To distinguish these two kinds of dimerization, we consider two rung tunneling amplitudes, $D_1 = \overline{\braket{c_{j,+}^{\dag} c_{j,-} + \text{h.c.}}}$ 
and $D_2 = \overline{\braket{c_{j,+}^{\dag} c_{j-1,-} + \text{h.c.}}}$, averaged over the site index $j$. The difference $D_1-D_2$, 
as seen in Fig.~\ref{fig:idmrg_orders}(a), can indeed characterize these two band insulator phases.

In Sec.~\ref{subsection:band-insulator}, we have shown by analyzing the non-local string and parity order parameters that the band insulator phase for $m < 0$ has a topological nature, while the same for $m > 0$ is trivial. Here, we numerically verify this analytical result by examining string ($\mathcal{O}_S$) and parity ($\mathcal{O}_P$) correlation functions  defined as~\cite{Hida1992, dalla-2006, Berg2008, Batrouni2013, Batrouni2014}:
\begin{align}
    \mathcal{O}_S &= \lim_{|i-j| \rightarrow \infty} \langle \hat{O}_S(i) \hat{O}_S(j)\rangle = 
    \lim_{|i-j| \rightarrow \infty} \langle \delta\hat{n}_i  e^{i \pi \sum_{l = i}^{j} \delta\hat{n}_l} \delta\hat{n}_j \rangle, \nonumber \\
    \mathcal{O}_P &= \lim_{|i-j| \rightarrow \infty} \langle \hat{O}_P(i) \hat{O}_P(j)\rangle =
    \lim_{|i-j| \rightarrow \infty} \langle   e^{i \pi \sum_{l = i}^{j} \delta\hat{n}_l}  \rangle,
    \label{eq:string_parity}
\end{align}
where the operators $\hat{O}_S$ and $\hat{O}_P$ are defined in Eqs.~\eqref{O-PS}.
with $\delta\hat{n}_j$ being the density fluctuation 
across the zigzag rung $j$. 
It is to be noted that these non-local order parameters can be measured experimentally in cold atomic setups~\cite{Endres2011, Hilker, Lesleuc2019, Sompet2022, Wei2023}. In Figs.~\ref{fig:idmrg_orders}(b) and (c), we show that for the topological band insulator ($m<0$) the string correlation is non-zero, while the parity correlation vanishes (see also Fig.~\ref{fig:idmrg_orders_cut}(a)) -- indicating hidden non-local order similar to topological Haldane insulators~\cite{dalla-2006, Berg2008, Batrouni2013, Batrouni2014}. The opposite is true for the trivial band insulator, i.e., $\mathcal{O}_S = 0, \mathcal{O}_P \neq 0$.

For further verification of the topological nature of the band insulator phases, we  consider the entanglement gap $\Delta \varepsilon = \varepsilon_1 - \varepsilon_0$, where $\varepsilon_0$ and $\varepsilon_1$ are the ground and first excited state energies of the entanglement Hamiltonian $H^{E}_l$ respectively. The entanglement Hamiltonian is defined as $H^E_l = -\ln \rho_l$, where $\rho_l$ is the $l$-site reduced density matrix. It has been established that the entanglement Hamiltonian possesses degenerate spectra for phases with topological properties in one dimension~\cite{Pollmann_prb_2010}. In Fig.~\ref{fig:idmrg_orders}(d) we plot the the entanglement gap $\Delta \varepsilon$ in the $(m/t_0, V/t_0)$-plane. Clearly, vanishing $\Delta \varepsilon$ in the band insulator phase for $m<0$ dictates the topological nature of this phase (see also Fig.~\ref{fig:idmrg_orders_cut}(b)).

For large values of $V/t_0$, we end up with a spontaneous symmetry-broken (SSB) phase where the $\mathbb{Z}_2$-symmetry corresponding to $\mathcal{P}$ gets spontaneously broken. As a result, the relative density $\rho_{\text{rel}} = \overline{\braket{\hat n_{j, +} - \hat n_{j, -}}}$ between two legs becomes non-zero (see Fig.~\ref{fig:idmrg_orders}(e)) and serves as an order parameter to detect this SSB phase, see Sec.~\ref{sec:subsubsec3}. As 
discussed in Sec.~\ref{sec:m0}, the transition from the Gaussian criticality to this SSB phase along $m=0$ line is Berezinskii-Kosterlitz-Thouless (BKT) type. This is why the order parameter $\rho_{\text{rel}}$ varies very smoothly along the $m=0$ line, as opposed to the case of $m \neq 0$ (Fig.~\ref{fig:idmrg_orders_cut}(b)) where the transitions from the band insulator phases to the SSB phase are of second order Ising type.

\subsection{Bifurcation of the criticality}
\label{subsec:bifur}

Now, we move to carefully analyze the splitting of the critical line at $m=0$ into two other critical lines at around $V/t_0 \simeq 2.45$ as seen in Fig.~\ref{fig:Ashkin-Teller-Phase-Diagram}(b). 

\begin{figure}
    \includegraphics[width=0.9\linewidth]{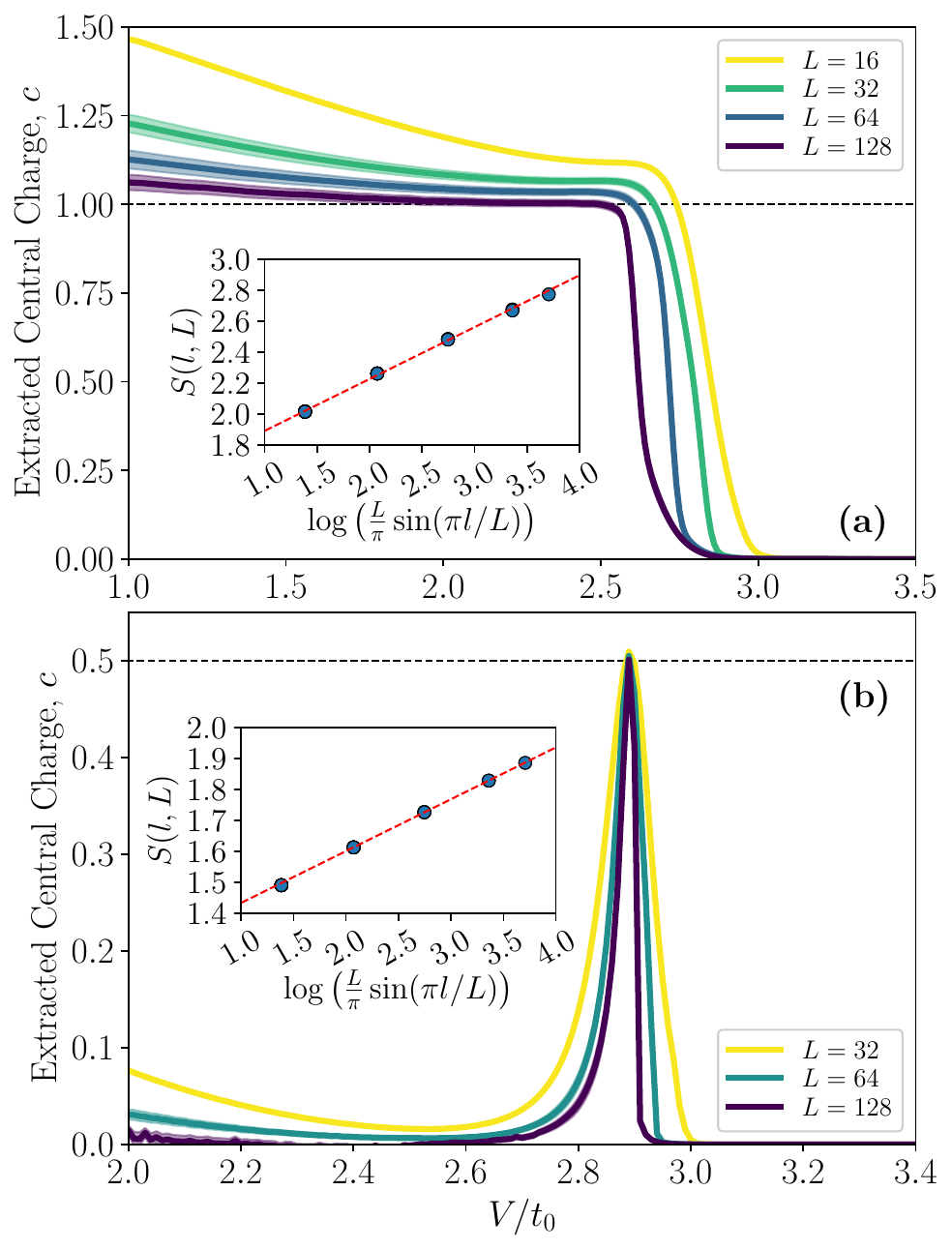}
    \caption{(Color online.) 
    The extracted values of the central charge for different system-sizes using the scaling function of Eq.~\eqref{eq:central_charge} for (a) $m=0$ and (b) $m=0.1 t_0$. The critical line at $m=0$ has central charge $c=1$, i.e., a U(1) Gaussian criticality, while the bifurcated critical lines  at $m \neq 0$ belong to the Ising universality class having the central charge $c=1/2$. The shaded regions mark the errors in the fitting procedure.
    (Insets) The fitting of the entanglement entropy according to Eq.~\eqref{eq:central_charge} for (a) $m=0$ and $V/t_0 = 2.3$ resulting in $c = 1.00(2)$, and (b) $m=0.1 t_0$ and $V/t_0 = 2.89$ resulting in $c=0.50(1)$. For the insets, we have chosen the data for $L=128$. 
    }
    \label{fig:ttn_m0_central_charge}
\end{figure}

For this purpose, first, we determine the central charges $c$ of the underlying conformal field theory (CFT) for these critical lines using the finite-bipartition scaling of von Neumann entanglement entropy. 
The von Neumann entanglement entropy of a block of $l$ sites is defined as 
\begin{equation}
    S(l) = - \text{Tr} \left[ \rho_l \ln(\rho_l) \right],
\end{equation}
where $\rho_l = \text{Tr}_{l+1, l+2, \cdots, L} \ket{\psi}\bra{\psi}$ is the $l$-site reduced density matrix after tracing out rest of the system. 
In a CFT, the finite-size scaling of the entanglement entropy of a bipartition of size $l$ in a system of length $L$ with PBC is~\cite{callan_geometric_1994, Vidal2003, calabrese_entanglement_2004}:
\begin{equation}
    S(l, L) = \frac{c}{3} \ln \left[\frac{L}{\pi} \sin(\pi l/L)  \right] + b',
    \label{eq:central_charge}
\end{equation}
where $b'$ is a non-universal constant. In Fig.~\ref{fig:tn_ansatz}, we show the variations of the fitted values of the central charge, according to Eq.~\eqref{eq:central_charge}, as functions of $V/t_0$ for $m=0$ (Fig.~\ref{fig:ttn_m0_central_charge}(a)) and $m=0.1$ (Fig.~\ref{fig:ttn_m0_central_charge}(b)). The figure clearly indicates that the critical line at $m=0$ has central charge $c=1$, and therefore describes $U(1)$ Gaussian criticality. On the other hand, two bifurcated critical lines at $m \neq 0$ has $c=1/2$ and thereby describes the Ising criticality. This scenario matches that of the AT model and our analytical prediction from the phenomenological analysis of the DSG model.

\begin{figure}
    \includegraphics[width=\linewidth]{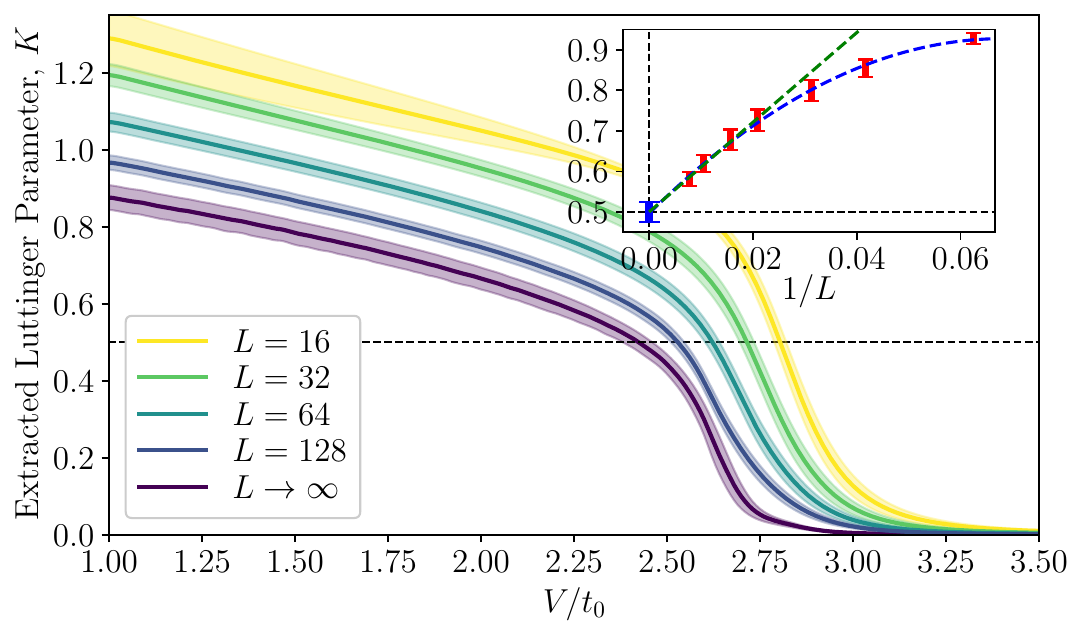}
    \caption{(Color online.)
    The extracted values of the Luttinger parameter $K$ for different system-sizes using the scaling form of Eq.~\eqref{eq:luttingerKscaling} along the $m=0$ line. The shaded regions denote the error-bars. The estimated values of $K$ in the thermodynamic limit has been extracted by using a linear function $f_1(1/L) = K_{\infty} + b/L$ and a quadratic function $f_2(1/L) = K_{\infty} + b/L + d/L^2$ in $1/L$ (see inset).}
    \label{fig:ttn_m0_luttingerK}
\end{figure}

However, the scaling of entanglement entropy does not shed much lights on the bifurcation point, and it is unable to tell us whether this is a SU(2)$_1$ critical point. To confirm that this bifurcation point is indeed a SU(2)$_1$ critical point, we determine the Luttinger parameter $K$ numerically and show that it tends to $1/2$ at the bifurcation point (see Sec.~\ref{sec:Ashkin-Teller}).

In this analysis, we extract the Luttinger parameter $K$ from the scaling of the bipartite fluctuations~\cite{Song10, Song12, Rachel12}.  
In a Luttinger liquid with a global $\text{U}(1)$ conserve quantity $O$ and with PBC, the local fluctuations
\begin{equation}
    \mathcal{F}_l (O) = \Braket{\left(\sum_{n\leq l} O_n \right)^2} - \Braket{\sum_{n \leq l} O_n}^2
\end{equation}
obey the scaling of the form~\cite{Song10, Song12, Rachel12}:
\begin{equation}
    \mathcal{F}_l (O) = \frac{K}{\pi^2} \ln \left[\frac{L}{\pi} \sin(\pi l/L)  \right] + \text{const.}
    \label{eq:luttingerKscaling}
\end{equation}
For our system, the global $\text{U}(1)$ conservation corresponds to the fermionic charge $O_n = \hat{n}_{n,+} + \hat{n}_{n,-}$. In Fig.~\ref{fig:ttn_m0_luttingerK}, we show the extracted Luttinger parameter for several system-sizes $L \in [16, 128]$ with PBC across the $m=0$ line. Interestingly, as we increase the system-size $L$, the point in $V/t_0$, where the fitted $K$ crosses the value $1/2$, approaches towards the expected bifurcation point $V/t_0 \simeq 2.45$. We extrapolate the Luttinger parameter $K$ in the thermodynamic limit by using both a linear function and a quadratic function in $1/L$ (see the inset of Fig.~\ref{fig:ttn_m0_luttingerK}). By this extrapolation,
we find that the Luttinger parameter $K$ becomes $1/2$ at $V/t_0 = 2.43(4)$.
This $K=1/2$ value of the Luttinger parameter
confirms the nature of the bifurcation point as the SU(2)$_1$ WZNW critical point in correspondence with the discussions of Sec.~\ref{sec:Ashkin-Teller}.

\section{Conclusion}
\label{sec:conclu}

In this work, we studied the phase diagram of a system of interacting spinless fermions on a two-leg triangular ladder at half-filling, with uniform $t_{0}$ intrachain and alternating $t_{1,2}$ interchain nearest-neighbor tunneling amplitudes, $f=1/2$ magnetic flux per triangular plaquette (in units of $\pi$), and $V$ nearest-neighbor density-density interchain interaction.
At the microscopic level, the model exhibits a U(1) symmetry pertained to the conservation of total fermion number and a $\mathbb{Z}_2$ symmetry - a combined parity transformation and the chain exchange operation.
The model is based on experimental setups with cold atom gases in optical lattices, in the presence of off-resonant laser driving to Rydberg states. The regime of parameters under consideration is in principle accessible experimentally \cite{Glaetzle2015,Dalmonte2015,jaksch2003creation,Guardado-Sanchez2021}.

To obtain the phase diagram, we use the bosonization approach in the weakly interacting regime ($|V| \ll t_{0}$), with $0<t_{1,2} \ll t_0$ and $|t_{1}-t_{2}| \ll t_{1}+t_{2}$, and map the model onto the double-frequency sine-Gordon model. We analytically predict various properties of the system, by utilizing the symmetries of the original lattice model, and renormalization group analysis for the bosonized version. Specifically, for $t_{1} \neq t_{2}$ and sufficiently weak repulsive interaction $V \ll V_{c}$, the system is a band insulator. If additionally $t_{1} > t_{2}$ is the case, then the phase is a trivial band insulator with non-zero dimerization along $t_{1}$ links. If $t_{1} < t_{2}$ holds, then the phase is instead a topological band insulator, with non-zero dimerization along $t_{2}$ links, and  displays edge states for open boundary conditions. For $t_{1} = t_{2}$, the system is described by a Gaussian model with central charge $c=1$, separating two band insulator phases.

The Gaussian critical line for $t_{1} = t_{2}$ terminates at $V=V_{c}$, where the symmetry of the system is enlarged from $\mathbb{Z}_{2} \times$U(1) to SU(2), with the underlying field theory of the model corresponding to SU(2)$_{1}$ Wess-Zumino-Novikov-Witten (WZNW) model. This emergent non-Abelian SU(2) invariance, that is absent in the microscopic description of the system, is a remarkable effect coming from the interplay between the geometric frustration, magnetic flux, and many-body correlations.

At this $t_{1} = t_{2}$ regime, when crossing $V=V_{c}$ critical point, the system undergoes a Berezinskii-Kosterlitz-Thouless  transition to a gapped phase, with spontaneously broken $\mathbb{Z}_{2}$ symmetry. In this phase, we observe non-zero charge imbalance (i.e., a net relative density between the two chains) and total current along the chains. Additionally, the Gaussian critical line, terminated at SU(2)$_{1}$ point, bifurcates into two Ising critical lines, with central charge $c=1/2$, similar to what is seen in the Ashkin-Teller model. These Ising critical lines separate the strong-coupling symmetry-broken phase from the band insulators.

Since the bosonization approach is valid for weak coupling regimes, we have used numerical simulations based on tensor network states to corroborate the analytical predictions. Specifically, using infinite density-matrix renormalization group (iDMRG) method, we characterize the phase diagram and different phases of the lattice Hamiltonian directly at the thermodynamic limit in the enlarged range of interaction strength $V$, and confirm the predictions of bosonization approach. By applying tree tensor network (TTN) based calculations for finite systems with periodic boundary conditions, we characterize both the Gaussian $c=1$ and Ising $c=1/2$ critical lines by using the scaling of entanglement entropies. Furthermore, from the numerical scaling of bipartite fluctuations corresponding to the global U(1) conserve quantity, we confirm the existence of SU(2)$_1$ WZNW bifurcation point similar to what is observed in the Ashkin-Teller model.

Our work, therefore, provides a unique example where non-Abelian SU(2) symmetry emerges in a fundamentally Abelian system.

\acknowledgements
We thank Poetri S. Tarabunga for precious discussions and collaborations during the implementations of the TTN codes. We are grateful to Simone Montangero, Simone Notarnicola, Pietro Silvi, and Colin Egan for the useful discussions regarding the developments of the code. M.D. thanks M. Fabrizio and P. Fendley for discussions. A.N. thanks F. H. L. Essler and O. Starykh for their interest in our work and useful comments. B.B. and A.N. acknowledge fruitful cooperation with G. Japaridze
on projects related to frustrated one-dimensional quantum systems.
T.C. acknowledges the support of PL-GRID infrastructure for the computational resource. 
M.T. thanks the Simons Foundation for supporting his Ph.D. studies through Award 284558FY19 to the ICTP.
The work of M.D. was partly supported by the ERC under grant number 758329 (AGEnTh), and by the Munich Institute for Astro-, Particle and BioPhysics (MIAPbP) which is funded by the Deutsche Forschungsgemeinschaft (DFG, German Research Foundation) under Germany´s Excellence Strategy – EXC-2094 – 390783311. The support of B.B and A.N. from the Shota Rustaveli National Science Foundation of Georgia, SRNSF, Grant No. FR-19-11872, is gratefully acknowledged. M.D. and E.T. further acknowledge support from the MIUR Programme FARE (MEPH), and from  QUANTERA DYNAMITE PCI2022-132919.
The iDMRG simulations have been performed with the TeNPy library~\cite{tenpy18}, while the TTN simulations use the C++ ITensor library~\cite{itensor22} as its backbone.

\appendix

\section{Continuum form of local physical operators in the chain and band representations} \label{appendix:ops}


In the continuum limit, after projecting  onto
the low-energy sector,  the fluctuation parts of 
local physical fields, defined in the chain representation, take the following form:

Local densities on each chain:
\begin{equation}
\label{den-blu}
    \begin{split}
        \hat{n}_{n\sigma}:
            &\equiv :c^{\dag}_{n,\sigma} c_{n,\sigma}:
            ~ \to ~ a_{0}\rho_{\sigma} (x), \\
        \rho_{\sigma} (x) 
            &=  :\Psi^{\dag}_{\sigma} (x) \Psi_{\sigma} (x): .
    \end{split}
\end{equation}

Longitudinal currents on each chain at $f=1/2$:
\begin{align}
\nonumber
    :J^{\sigma}_{n,n+1}:
    = - & \mathrm{i}t_0
        \left(
            :c^{\dag}_{n,\sigma} c_{n+1,\sigma}:e^{-i\pi \sigma f } -
            \text{h.c.}
        \right)\Big|_{f=1/2}\\
\nonumber
    = - & t_0 \sigma
        \left(
            :c^{\dag}_{n,\sigma} c_{n+1,\sigma}: + \text{h.c.} 
        \right)\\
\nonumber
    \to & \sigma v_{\text{F}}  :\Psi^{\dag}_{\sigma} (x) \Psi_{\sigma} (x): + ~O(v_{\text{F}} a_{0}) \equiv j^0 _{\sigma} (x),
\end{align}
implying that, due to their chiral nature, in the leading order  at $a_{0} \to 0$, local densities and longitudinal currents coincide up to a prefactor $v_{\text{F}}$:
\begin{equation}
\label{long-curr-blu}
     j^{0}_{\sigma} (x) = \sigma v_{\text{F}} \rho_{\sigma} (x).
\end{equation}

Interchain currents on $t_{1}$ and $t_{2}$ zigzag links:
\begin{align*}
    :J^{+-}_{nn}:
        &\equiv -\mathrm{i}t_{1} :
        \left(c^{\dag}_{n,+} c_{n,-} - \text{h.c.} \right): \\
        &\to  t_{1} a_{0} \Psi^{\dag} (x) \hat{\sigma}_2 \Psi(x), \\
    :J^{-+}_{n,n+1}:
        &\equiv - \mathrm{i}t_{2} :
        \left(c^{\dag}_{n,-} c_{n+1,+} - \text{h.c.} \right): \\
        &\to ~ t_{2} a_{0} \Psi^{\dag} (x) \hat{\sigma}_2 \Psi(x).
\end{align*}
At $t_{1} = t_{2}$ ($m=0$) and $a_{0} \to 0$ with $v_{\text{F}} = 2t_0 a_{0} = \text{const}$, the currents along the oriented $t_{1}$ and $t_{2}$ links coincide:
\begin{equation}
\label{zig-curr-blu}
    \begin{split}
        :J^{+-}_{nn}: \to j_z (x),\quad:J^{-+}_{n,n+1}: \to  j_z (x),\\
        j_z (x) = v_{\text{F}} \tau  :\Psi^{\dag} (x) \hat{\sigma}_2 \Psi(x) : .
    \end{split}
\end{equation}

Bond-density fields:
\begin{equation}
\label{eq:B-field-blu}
     \begin{split}
        B_{nn} &= :c^{\dag}_{n,+} c_{n,-}: + \text{h.c.}
            \to a_{0} B (x), \\
        B_{n,n+1} &= :c^{\dag}_{n,+} c_{n-1,-}: + \text{h.c.}
            \to - a_{0} B (x),\\
        B(x) &=  :\Psi^{\dag} (x) \hat{\sigma}_{1} \Psi (x): .
     \end{split}
\end{equation}
In formulas (\ref{den-blu}-\ref{eq:B-field-blu}), normal ordering prescription is defined as $:\hat{A}: = \hat{A} - \langle \hat{A} \rangle_{0}$, where averaging is done over the vacuum of the noninteracting model at $f=1/2$ and $m=0$. From formulas (\ref{den-blu}) and (\ref{long-curr-blu}) it follows that the total current $j_{0} = \sum_{\sigma} j_{0}^{\sigma}$ of the zigzag ladder at $f=1/2$ is proportional to the relative particle density
\begin{equation}
j_0 (x) = v_{\text{F}} \rho_\text{rel} (x) \label{tot-curr-rel-den}.
\end{equation}

Using the transformations \eqref{eq:u-transf} and passing to the rotated basis, we obtain the expressions for all above operators in the band representation:
\begin{align}
\nonumber
    \rho_{+} (x)&= \frac{1}{v_{\text{F}}} j_{0} ^+ (x) \\
\label{eq:ro+rot}
        &= u^{2} J_{\text{R}} (x) + v^{2} J_{\text{L}} (x) -
        uv\mathcal{N}_2(x),\\
\nonumber
    \rho_{-} (x) &= - \frac{1}{v_{\text{F}}} j_{0} ^- (x) \\
\label{ro-rot}
        &= v^{2} J_{\text{R}} (x) + u^{2} J_{\text{L}} (x) +
        uv\mathcal{N}_2(x),\\
\nonumber
    j_z (x) &= v_{\text{F}} \tau
        \big[
            2uv (J_{\text{R}} (x) - J_{\text{L}} (x) ), \\
\label{zig-curr-rot}
            &\hspace{23mm}
            + (u^{2} -  v^{2})\mathcal{N}_2(x)
        \big],\\
\label{eq:D-rot}
    B(x) &= \mathcal{N}_1 (x).
\end{align}
Here $J_{\text{R}} (x) = :R^{\dag}(x)R(x):$ and $J_{\text{L}} (x) = :L^{\dag} (x)L (x): $ are U(1) chiral fermionic currents defined in the band basis (see e.g., Ref. \cite{Nersesyan2004}), and $ \mathcal{N}_{1,2} $
are Dirac mass bilinears:
\begin{align}
\nonumber
    \mathcal{N}_1(x)
        &= \chi^{\dag}(x)\hat{\sigma}_{1} \chi (x) \\
\label{N1}
        &= :R^{\dag} (x)L (x):+ :L^{\dag} (x)R (x):, \\
\nonumber
    \mathcal{N}_2(x)
        &= \chi^{\dag}(x) \hat{\sigma}_{2} \chi  (x) \\
\label{N2}
        &= - \mathrm{i}
        \left[ :R^{\dag} (x)L (x):- :L^{\dag} (x)R (x): \right].
\end{align}
The expressions (\ref{zig-curr-rot}), (\ref{eq:D-rot}) are the leading terms of the expansion in small $a_{0}$. Under the $\mathcal{P}$-transformation
\begin{eqnarray}
R(x) \to L(-x),  && L(x) \to  R(-x), \nonumber\\
 J_{\text{R}} (x) \to J_{\text{L}} (-x), && J_{\text{L}} (x) \to J_P (-x),\nonumber\\
 \mathcal{N}_1 (x) \to \mathcal{N}_1 (-x), &&\mathcal{N}_2 (x) \to - \mathcal{N}_2 (-x). \label{eq:P-transf}
\end{eqnarray}

Here a comment is in order. In models of 1D lattice fermions with a half-filled band, operators with the structure (\ref{N1}), (\ref{N2}) 
are associated with spatially modulated (staggered) order parameter fields. In those cases, the fermionic bilinears $R^{\dag}L, ~L^{\dag}R$ emerge due to  hybridization of single-particle states near two {opposite} Fermi points, with the momentum transfer close to $2k_{\text{F}} = \pi$. In the present model, there is {only one Dirac point} in the low-energy spectrum, and the particle-hole fields with momentum transfer $\pi$ are all short-ranged. In fact, the appearance of the fermionic bilinear $\mathcal{N}_{2}$  in the asymptotic expressions  (\ref{eq:ro+rot})--(\ref{zig-curr-rot}) is entirely due to the $\tau$-deformation of the kinetic energy (\ref{eq:psi}), that is geometrical frustration of the zigzag ladder.

\section{Boundary modes in the topological phase of band insulator} \label{appendix:boundary}

In Sec.~\ref{subsection:band-insulator} of the main text we have shown that under the conditions $K>1/2$ and $m\neq 0$, for both signs of the ``light'' mass $m$ the ladder displays a band insulator phase with massive Dirac fermions being elementary low-energy excitations. In this Appendix, we  address the topological properties of these phases by studying zero-energy boundary  states in a semi-infinite sample of a triangular 1/2-filled flux ladder at $f=1/2$. Since (apart from a possible formation of excitonic states) at $K>1/2$ interaction effects basically reduce to renormalization of the single-particle mass gap, Eq.~\eqref{gen-mass}, it is sufficient to do the calculation for a noninteracting model.
\medskip

Consider a semi-infinite sample, in which the diatomic unit cells are labeled as $n=1,2, \ldots , \infty$. Let us adopt the continuum limit of this model by taking into account both Dirac-like low-energy modes  with masses $M = t_{1}+t_{2}$ and $m=t_{1} - t_{2}$, $|M|,|m| \ll t_0$. Then we can write
\begin{equation}
    c_{n,\sigma} \to \sqrt{a_{0}}
        \left[
            (-1)^n \psi_{\sigma} (x) + \bar{\psi}_{-\sigma} (x)
        \right],
    \quad (\sigma=\pm),
\end{equation}
where $\psi_{\sigma} (x)$ and $\bar{\psi}_{\sigma} (x)$ are fermionic fields describing single-particle excitations with momenta close to $\pi$ and $0$, respectively. Adding an additional rung $n=0$ to the open end of the ladder we impose boundary conditions
\begin{equation}
\label{eq:bc}
    c_{0,\sigma} = 0 ~\to ~ 
        \psi_{\sigma} (0) + \bar{\psi}_{-\sigma} (0) =0,
    \quad
    (\sigma=\pm).
\end{equation}
Denoting by $\{u(x),v(x)\}$ and $\{\tilde{u}(x), \tilde{v}(x)\}$ the components of the 2-spinor wave functions $w(x)$ and $\tilde{w}(x)$
associated with the field operators $\psi(x)$ and $\bar{\psi}(x)$, from \eqref{eq:bc} we obtain
\begin{equation}
    u(0) + \tilde{v}(0) = 0, \quad
    v(0) + \tilde{u} (0) = 0. \nonumber
\end{equation}
This leads to the following constraint imposed on the boundary spinors:
\begin{equation}
\label{boundary-spinors}
    w(0) =
        \begin{pmatrix}
        u(0)\\
        v(0)
        \end{pmatrix},
    \quad
    \tilde{w}(0) =
        \begin{pmatrix}
        \tilde{u}(0)\\
        \tilde{v}(0)
        \end{pmatrix}
    = - \hat{\sigma}_{1} w(0). 
\end{equation}
The boundary zero modes corresponding to these functions satisfy
the equations
\begin{align}
\label{eq:EQ1}
    \big[
        - \mathrm{i}v_{\text{F}}
        \left( \sigma_{3} + \tau \sigma_{2} \right) - m \sigma_{1}
    \big]
    w(x) &= 0,\\
\label{eq:EQ2}
    \big[
        - \mathrm{i}v_{\text{F}}
        \left( \sigma_{3} + \tau \sigma_{2} \right) - M \sigma_{1}
    \big]
    \tilde{w}(x) &= 0, ~~~x \geq 0.
\end{align}
The  kinetic energy in  Eqs. \eqref{eq:EQ1} and \eqref{eq:EQ2} is diagonalized by an SU(2) transformation of the spinors 
\begin{eqnarray}
\label{U1}
    \psi = U\zeta,\quad
    \bar{\psi} = U \tilde{\zeta},
\end{eqnarray}
where
\begin{eqnarray}
    \zeta =
        \begin{pmatrix}
        z_1\\
        z_2
        \end{pmatrix}, &&
    \tilde{\zeta}=
        \begin{pmatrix}
        \tilde{z}_{1}\\
        \tilde{z}_{2}
        \end{pmatrix},
\end{eqnarray}
and the unitary matrix $U$ is defined in \eqref{eq:u-transf}.
The spinors $\zeta$ and $\tilde{\zeta}$ satisfy canonical Dirac equations for zero modes:
\begin{align*}
    \left( - \mathrm{i}\tilde{v} \sigma_{3} \partial_{x} - m \sigma_{1} \right) \zeta(x) &= 0,\\
    \left( - \mathrm{i}\tilde{v} \sigma_{3} \partial_{x} - M \sigma_{1} \right) \tilde{\zeta}(x) &= 0.
\end{align*}
On the semi-axis $x \geq 0$ their solution reads
\begin{equation}
\label{dir-solutions-1}
    \begin{split}
        \zeta(x) &= \zeta_{0} 
            \begin{pmatrix}
            1\\
            \mathrm{i}s_m
            \end{pmatrix}
            \exp \left( - |m|x/\tilde{v} \right),
        \\
        \tilde{\zeta}(x) &= \tilde{\zeta}_{0} 
            \begin{pmatrix}
            1\\
            \mathrm{i}s_M
            \end{pmatrix}
            \exp \left( - |M|x/\tilde{v} \right),
    \end{split}
\end{equation}
where $\zeta_{0}$ and $\tilde{\zeta}_{0}$ are normalization coefficients and $s_m = \text{sgn}~m$, $s_M = \text{sgn}~M$.
\medskip

Using the transformations \eqref{U1} we obtain
\begin{align}
\label{psi-x}
    w(x) &=
        \chi_0
        \begin{pmatrix}
            u - v s_m \\
            \mathrm{i}(v + u s_m)
        \end{pmatrix}
        \exp \left( - |m|x/\tilde{v}  \right),\\
\label{tpsi-x}
    \tilde{w}(x) &=
        \tilde{\chi}_0
        \begin{pmatrix}
            u - v s_M \\
            \mathrm{i}(v + u s_M)
        \end{pmatrix}
        \exp \left( - |M|x/\tilde{v}  \right).
\end{align}
On the other hand, according to the boundary condition \eqref{boundary-spinors}: 
\begin{equation}
\label{extra}
    \tilde{w}(x) = - \chi_0
        \begin{pmatrix}
            \mathrm{i}(v + u s_m)\\
            u - v s_m
        \end{pmatrix}
        \exp \left( - |M|x/\tilde{v}  \right).
\end{equation}
Then we obtain
\begin{align*}
    \tilde{\zeta}_{0} (u - v s_M)
        &= - \mathrm{i}\zeta_{0} (v + u s_m), \\
    \mathrm{i}\tilde{\zeta}_{0} (v+u s_M)
        &= - \zeta_{0} (u - v s_m),
\end{align*}
implying that
\begin{eqnarray}
    \frac{\tilde{\zeta}_{0}}{\zeta_{0}}
        = \frac{- \mathrm{i}(v+us_m)}{u - v s_M}
        = \frac{\mathrm{i}(u-vs_m)}{v+u s_M}.
\end{eqnarray}
From the last equation in follows that
\begin{eqnarray}
&(u - vs_m)(u - v s_M)  + 
(v + u s_m)(v + u s_M)
\nonumber \\
&= 1 +   s_m s_M = 0, 
\end{eqnarray}
which leads to the conclusion that a normalizable boundary zero mode only exists --  and hence, according to the bulk-boundary theorem,
the ground state is topologically nontrivial --
if the masses $m$ and $M$
have different signs:
\begin{equation}
s_m s_M = -1 ~~~~\rightarrow ~~~~ Mm < 0.
\label{result}
\end{equation}
With the convention $M>0$ adopted in the main text, the ground state  at $K>1/2$ represents a topological insulator if $m<0$
and is topologically trivial at $m>0$.  

The total wave function for the boundary zero mode has the structure
\begin{eqnarray}
\Upsilon (x) = (-1)^{x/a_{0}} w(x) + \tilde{w}(x), ~~~x\geq 0,
\label{total-psi}
\end{eqnarray}
where $w(x)$ and $\tilde{w}(x)$ are given by expressions \eqref{psi-x} and \eqref{tpsi-x}, respectively, in which
the condition \eqref{result} has to be taken into account. If $|M| \gg |m|$, then
$\tilde{w}(x)$ exponentially decays at short distances, $x \sim \tilde{v}/|M|$. At longer distances, $x \gtrsim \tilde{v}/|m|$,
there exists an exponential tail of  the boundary wave function contributed by the light fermions.

\begin{figure}[t]
    \includegraphics[width=\linewidth]{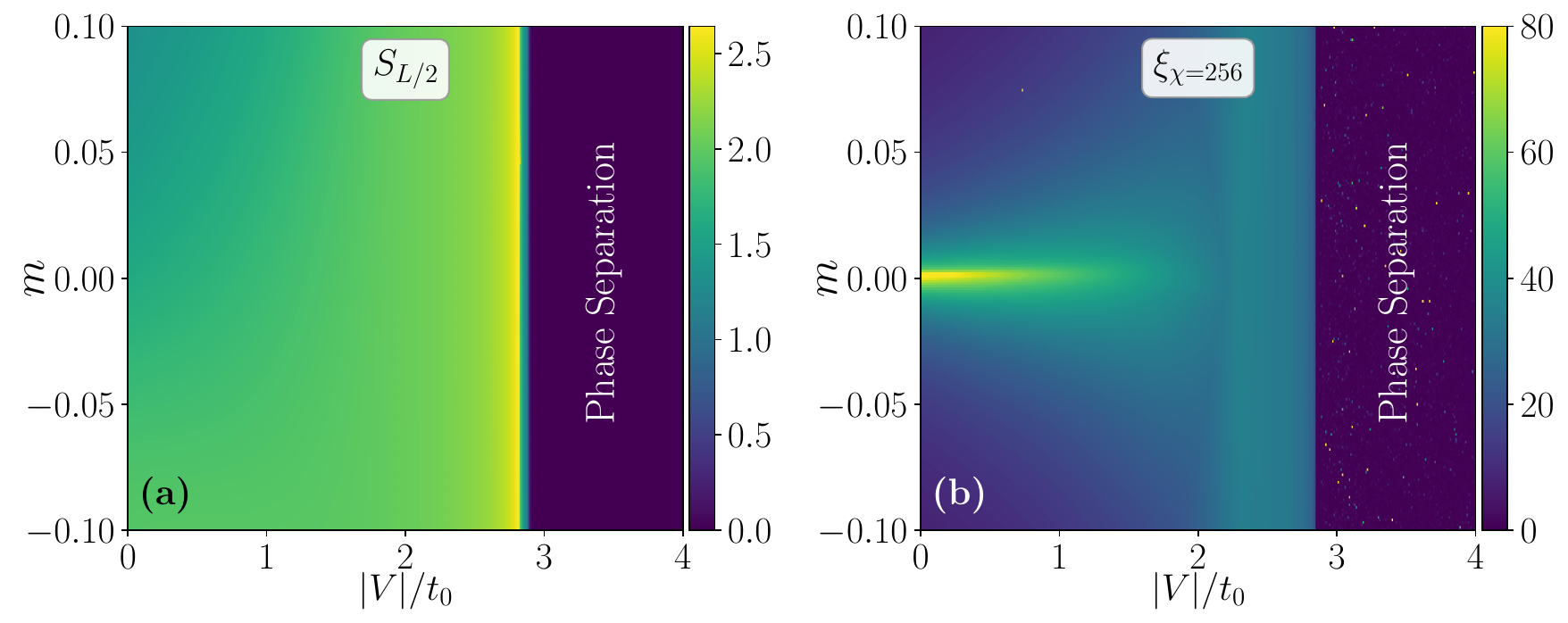}
    \caption{(Color online.)
    The phase diagram for an attractive triangular ladder in the $(m, V/t_0 < 0)$-plane. (a) We plot half-system entanglement entropy for a finite ladder of linear size $L=60$ with OBC using fDMRG.
    (b) The correlation length $\xi_{\chi}$ extracted from iDMRG simulations with iMPS bond dimension $\chi=256$.
    }
    \label{fig:idmrg_phase_neg}
\end{figure}

\section{Numerical results for the attractive interchain interaction}
\label{appendix:attrac}

For the sake of completeness, we  consider the attractive interaction, i.e., $V < 0$ (thus $\lambda<0$ from Eq.~\eqref{eq:couplings}), between the chains.

From our analysis using bosonization, we have predicted that the scenario of attractive interaction is less interesting compared to the repulsive case. For the attractive regime, we have $\lambda<0$ and thus $K>1$. In this case, $\lambda$-term in Eq.~\eqref{eq:H-dsg}, which describes interband pair-hopping processes, is irrelevant. This way the properties of the model are determined only by the single-particle mass perturbation. As long as the attractive interaction is weak, where the mass term is relevant, we remain in either trivial or topological band insulator phases, depending on the sign of $m$. For very strong interaction $|V|/t_0 \gg 1$, the mass term becomes irrelevant, and the system phase separates between density $\rho=1$ Mott phase and density $\rho=0$ vacuum state. Since these phase separation states break the transnational invariance over macroscopic distances, iDMRG is not suitable for these states and randomly gets stuck to higher energy states. That is why we also employ finite DMRG (fDMRG) with OBC along with iDMRG simulations to confirm our results.

In Figs.~\ref{fig:idmrg_phase_neg}(a) and (b), we show the half-system entanglement entropy for finite ladder with OBC and the correlation length $\xi_{\chi}$ extracted from iDMRG simulations, respectively. Clearly, apart from the appearance of the phase separation region, the situation here is not that interesting unlike the situation of repulsive interactions.

\bibliographystyle{apsrev4-1}
\bibliography{revised.bbl}

\end{document}